\documentclass[aps,nofootinbib,floatfix,showpacs,preprintnumbers]{revtex4} 
\usepackage{graphicx}
\input epsf

\def\d{{\rm d}}

\def\lsim{\mathrel{\raise.3ex\hbox{$<$\kern-.75em\lower1ex\hbox{$\sim$}}}}
\def\gsim{\mathrel{\raise.3ex\hbox{$>$\kern-.75em\lower1ex\hbox{$\sim$}}}}

\def\cmm2{{\,\rm cm^{-2}}}
\def\cm2{{\,{\rm cm}^2}}
\def\cmm3{{\,{\rm cm}^{-3}}}
\def\gcmm3{{\,{\rm g\,cm^{-3}}}}

\def\fun#1#2{\lower3.6pt\vbox{\baselineskip0pt\lineskip.9pt
  \ialign{$\mathsurround=0pt#1\hfil##\hfil$\crcr#2\crcr\sim\crcr}}}

\def\be{\begin{equation}}
\def\ee{\end{equation}}
\def\bea{\begin{eqnarray}}
\def\eea{\end{eqnarray}}

\def\sigv{\langle\sigma v\rangle}

\begin{document}

\title{Dark Matter Annihilation in The Galactic Center As Seen by the Fermi Gamma Ray Space Telescope}

\author{Dan Hooper$^{1,2}$ and Lisa Goodenough$^3$}

\affiliation{$^1$Center for Particle Astrophysics, Fermi National
Accelerator Laboratory, Batavia, IL~~60510}
\affiliation{$^2$Department of Astronomy \& Astrophysics, The
University of Chicago, Chicago, IL~~60637}
\affiliation{$^3$Center for Cosmology and Particle Physics, Department of Physics, New York University, New York, NY~~10003}

\date{\today}
\begin{abstract}

We analyze the first two years of data from the Fermi Gamma Ray Space Telescope from the direction of the inner 10$^{\circ}$ around the Galactic Center with the intention of constraining, or finding evidence of, annihilating dark matter. We find that the morphology and spectrum of the emission between 1.25$^{\circ}$ and 10$^{\circ}$ from the Galactic Center is well described by a the processes of decaying pions produced in cosmic ray collisions with gas, and the inverse Compton scattering of cosmic ray electrons in both the disk and bulge of the Inner Galaxy, along with gamma rays from known points sources in the region. The observed spectrum and morphology of the emission within approximately 1.25$^{\circ}$ ($\sim 175$ parsecs) of the Galactic Center, in contrast, departs from the expectations for by these processes. Instead, we find an additional component of gamma ray emission that is highly concentrated around the Galactic Center. The observed morphology of this component is consistent with that predicted from annihilating dark matter with a cusped (and possibly adiabatically contracted) halo distribution ($\rho \propto r^{-\gamma}$, with $\gamma=1.18$ to 1.33). The observed spectrum of this component, which peaks at energies between 1-4 GeV (in $E^2$ units), can be well fit by a 7-10 GeV dark matter particle annihilating primarily to tau leptons with a cross section in the range of $\sigv = 4.6 \times 10^{-27}$ to $5.3 \times 10^{-26}$ cm$^3$/s, depending on how the dark matter distribution is normalized. We also discuss other sources for this emission, including the possibility that much of it originates from the Milky Way's supermassive black hole.

\end{abstract}
\preprint{FERMILAB-PUB-10-414-A}
\maketitle

\section{Introduction}

The inner volume of the Milky Way is one of the most promising targets for the indirect detection of dark matter. In particular, due to the high densities of dark matter predicted to be present in the region, the innermost tens or hundreds of parsecs of our galaxy is generally expected to produce the single brightest source of dark matter annihilation products, including gamma rays~\cite{gc}. 

Since its launch in June of 2008, the Large Area Telescope (LAT) onboard the Fermi Gamma Ray Space Telescope (FGST) has been observing gamma rays in the range of approximately 300 MeV to 100 GeV over the entire sky, including from the direction of the Galactic Center. Data from the FGST has been used to search for dark matter annihilations taking place in dwarf spheroidal galaxies~\cite{dwarfs}, galaxy clusters~\cite{clusters}, over cosmological volumes~\cite{cosmo}, throughout the Galactic Halo~\cite{galdif}, in dark matter subhalos~\cite{matt}, and in the Galactic Center~\cite{lisa}. In this article, we revisit the Galactic Center region as observed by the FGST and perform a detailed study of the spectral and morphological features of these gamma rays with the intention of either placing constraints on, or finding evidence for, dark matter annihilations taking place. In the course of this study, we have identified evidence for a component of emission that is highly concentrated in the inner degree around the Galactic Center, and which is spectrally and morphologically consistent with the spectrum and angular distribution predicted from annihilating dark matter in a cusped, and possibly adiabatically contracted, halo profile ($\rho \propto r^{-\gamma}$, with $\gamma=1.18$ to 1.33), with a mass of 7-10 GeV, and annihilating primarily to tau leptons. The normalization of the observed component requires the dark matter annihilation cross section to fall within the range of  $\sigv = 4.6 \times 10^{-27}$ to $5.3 \times 10^{-26}$ cm$^3$/s, in agreement with that required of dark matter in the form of a simple thermal relic.

The remainder of this paper is structured as follows. In the following section, we discuss the process of dark matter annihilation and the calculation of the angular distribution and spectrum of gamma ray annihilation products from the Galactic Center. In Sec.~\ref{bg}, we study the gamma ray emission as observed by the FGST in the inner 2-10$^{\circ}$ around the Galactic Center, and find it to be well described by pion decay and inverse Compton scattering in the disk and bulge of the Inner Galaxy, in addition to containing emission from known point sources. In Sec.~\ref{inner}, we study in the inner 2$^{\circ}$ and identify a component of emission highly concentrated around the Galactic Center. In Sec.~\ref{dm}, we discuss this component within the context of annihilating dark matter. In Sec.~\ref{other}, we discuss other possibilities for the origin of this component, such as unresolved point sources or the decays of energetic pions, but argue that these are unlikely to account for the observed emission. The most plausible astrophysical explanation for the observed emission is the Milky Way's supermassive black hole, although this explanation requires strong deviations from the power-law spectrum observed at higher energies. In Sec.~\ref{previous}, we compare the results of this study to those presented in our previous work~\cite{lisa}. Finally, in Sec.~\ref{discussion}, we discuss some of the implications of these results and draw our conclusions.

\section{Gamma Rays From Dark Matter Annihilations Near The Galactic Center}\label{darkmatter}

Dark matter halos are most dense at their centers. If the dark matter consists of particles that can self-annihilate, the innermost volume of the Milky Way's halo potentially may produce an observable flux of dark matter annihilation products, including gamma rays. The energy and angular dependent flux of such gamma rays is described by
\begin{equation}
\Phi_{\gamma}(E_{\gamma},\psi) =  \frac{\d N_{\gamma}}{\d E_{\gamma}} \frac{\sigv}{8\pi m^2_X} \int_{\rm{los}} \rho^2(r) \d l,
\label{flux1}
\end{equation}
where $\sigv$ is the dark matter annihilation cross section multiplied by the relative velocity of the two dark matter particles (averaged over the velocity distribution), $m_X$ is the mass of the dark matter, $\psi$ is the angle observed relative to the direction of the Galactic Center, $\rho(r)$ is the dark matter density as a function of distance to the Galactic Center, and the integral is performed over the line-of-sight. $\d N_{\gamma}/\d E_{\gamma}$ is the gamma ray spectrum generated per annihilation, which depends on the mass and dominant annihilation channels of the dark matter particle.

It is convenient to rewrite this equation as 

\begin{eqnarray}
\Phi_{\gamma}(E_{\gamma},\psi) \approx 2.8 \times 10^{-10} \, {\rm cm}^{-2} \, {\rm s}^{-1}\, {\rm sr}^{-1} \times \frac{dN_{\gamma}}{dE_{\gamma}}\, \bigg(\frac{\sigv}{3 \times 10^{-26} \,\rm{cm}^3/\rm{s}}\bigg) \, \bigg(\frac{100 \, \rm{GeV}}{m_{\rm{X}}}\bigg)^2 \, J(\psi) \,
\label{flux2}
\end{eqnarray}
where the dimensionless function $J(\psi)$ depends only on the dark matter distribution and is defined by convention as
\begin{equation}
J(\psi) = \frac{1}{8.5 \, \rm{kpc}} \bigg(\frac{1}{0.3 \, \rm{GeV}/\rm{cm}^3}\bigg)^2 \, \int_{\rm{los}} \rho^2(r(l,\psi)) {\rm d} l\,.
\label{jpsi}
\end{equation}
To calculate $J(\psi)$, a model for the dark matter halo distribution must be adopted. A commonly used parametrization of halo profiles is given by 
\begin{equation}
\rho(r) = \frac{\rho_0}{(r/R)^{\gamma} [1 + (r/R)^{\alpha}]^{(\beta - \gamma)/\alpha}} \,,
\label{profile}
\end{equation}
where $R \sim 20$-$30$ kpc is the scale radius of the halo and $\rho_0$ normalizes the halo's total mass. Among the most frequently used models is the Navarro-Frenk-White (NFW) profile, which is described by $\alpha = 1$, $\beta=3$ and $\gamma =1 $~\cite{nfw}. When considering the region of the Galactic Center, the most important feature of the halo profile is the inner slope, $\gamma$. In the inner parsecs of our galaxy, in which we are most interested, there are reasons to expect this slope to become steeper than described by NFW. In particular, the gravitational potential in the inner Milky Way is dominated by baryons rather than dark matter, which are not included in N-body simulations. Although the precise impact of baryons on the dark matter distribution is difficult to predict, an enhancement of the inner slope due to adiabatic contraction is generally expected~\cite{ac}.

The remaining factors in Eq.~(\ref{flux2}) depend on the properties of the dark matter particle, including its mass, annihilation cross section, and dominant annihilation channels.  A reasonable benchmark value for the annihilation cross section is $\sigv \approx 3 \times 10^{-26} \,\rm{cm}^3/\rm{s}$, which for a simple thermal relic will yield a density of dark matter that is similar to the cosmologically measured abundance~\cite{wmap}. If dark matter annihilations in the early universe take place largely through $P$-wave processes, resonant processes, or co-annihilations, then smaller values of  $\sigv$ are possible.

The challenge in identifying gamma rays from dark matter annihilations is not in merely observing the signal. A 50 GeV dark matter particle, for example, with $\sigv \approx 3 \times 10^{-26} \,\rm{cm}^3/\rm{s}$, and distributed according to NFW, should lead to the detection by the FGST of on the order of hundreds of gamma rays per year from dark matter in the inner degree of our galaxy. In order to make use of this signal, however, it must be distinguished from the various astrophysical backgrounds present in the Inner Galaxy~\cite{method}. Fortunately, the products of dark matter annihilations are predicted to possess distinctive characteristics that can be used to accomplish this. In particular, by searching for a highly concentrated signal around the Inner Galaxy (more centrally concentrated than diffuse astrophysical mechanisms, but more extended than a single point source), with the spectral shape consistent with that predicted from dark matter annihilations, it may be possible to identify a component of gamma rays originating from dark matter. In the following sections of this paper, we attempt to do this, by first turning our attention to the astrophysical gamma ray backgrounds in the region of the inner Milky Way.

\section{Modeling the Backgrounds of the Inner Galaxy}
\label{bg}

\begin{figure}[t]
\resizebox{10.0cm}{!}{\includegraphics{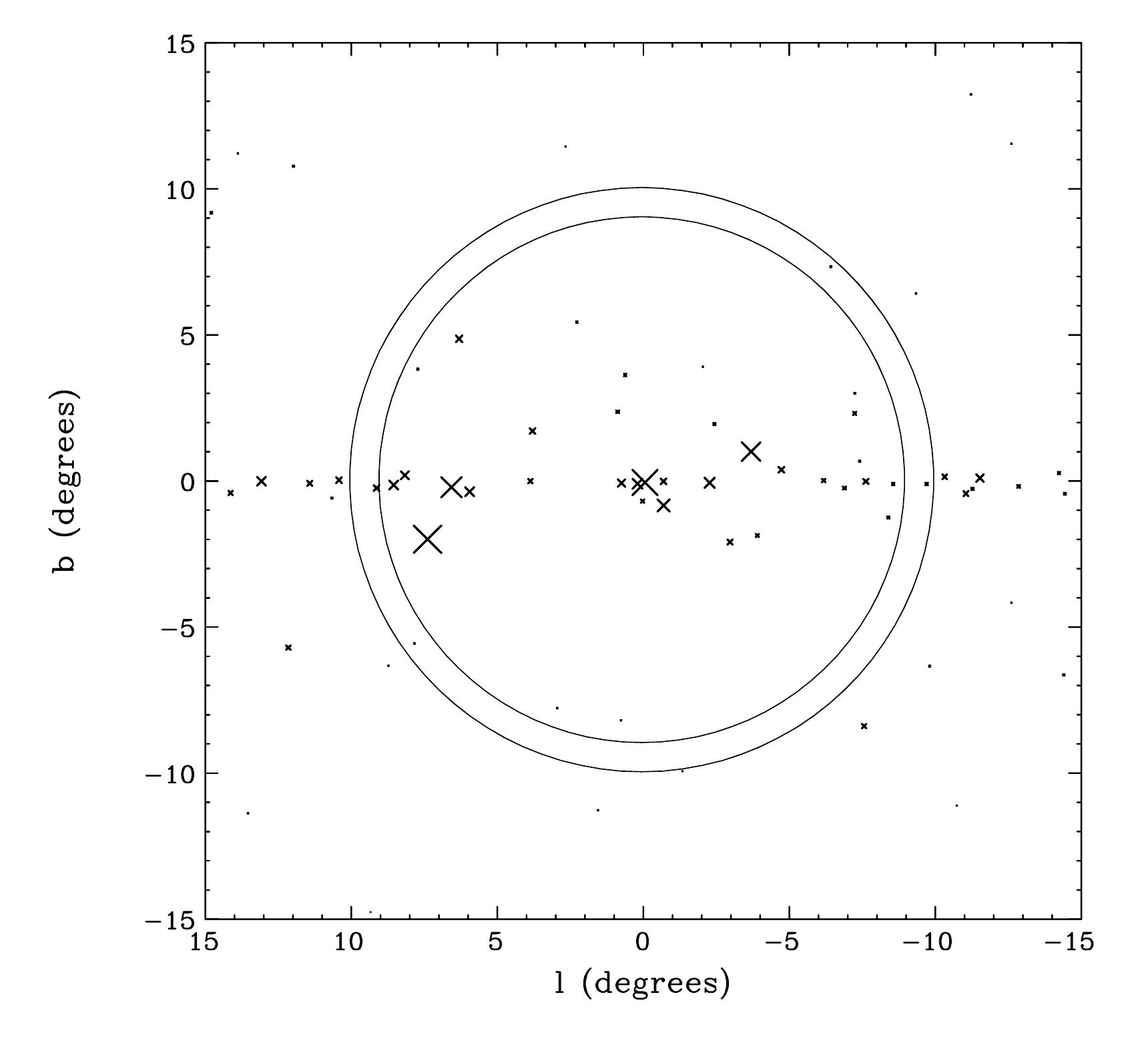}} \\
\caption{The locations of the 69 known point sources included in our study. The size of each $X$ is proportional to the intensity of that source in the range of 1-100 GeV, as described in the Fermi First Source Catalog. The region between the two solid circles is that shown in Figs.~\ref{angular1}-\ref{angular4}.}
\label{ptsources}
\end{figure}

We begin by modeling the emission from the region of the Galactic Center with three distinct components:
\begin{itemize}
\item{Emission from the Disk -- We model this component with a gaussian width around $b=0$, with the width, intensity, and spectral shape fit to the data. The width, intensity, and spectral shape of this component is allowed to vary freely with galactic longitude, $l$, although we find that these parameters do not significantly vary along the disk.}
\item{Emission distributed with spherical symmetry about the Galactic Center -- We model this component to have intensity and spectral shape that is constant with direction from the Galactic Center, but that can vary freely with distance to the Galactic Center.\footnote{Here and elsewhere throughout this study, the distance to the Galactic Center refers to distance to the dynamical center, which we take to be $b=0.0442^{\circ}$, $l=-0.055^{\circ}$~\cite{hess}, rather than $b=0$, $l=0$.} It is in this component that any contribution from dark matter annihilation would be expected to appear. Astrophysical contributions, such as those from processes taking place throughout the Galactic Bulge, also appear in this component.}
\item{Emission from known point sources -- We include the 69 sources in the Fermi First Source Catalog~\cite{catalog} with coordinates in the range of $l=(-15^{\circ},15^{\circ})$ and $b=(-15^{\circ}, 15^{\circ})$. We show the locations of these sources in Fig.~\ref{ptsources}, where the size of each $X$ is proportional to the intensity of the source over the energy range of 1-100 GeV.}
\end{itemize}

For a given set of model parameters (disk widths, spectral shapes, etc.), we convolve the emission with the the FGST's LAT point spread function (PSF), which describes the accuracy with which the arrival direction of a gamma ray can be determined. For front-converting (thin) events (due to their inferior PSF, we do not consider back-converting events in this study), the observed gamma ray flux (per solid angle) from a point source can be modelled according to\footnote{In the first version of this preprint, we had misinterpreted the definition of the Fermi PSF. We would like to acknowledge A.~Boyarksy, D.~Finkbeiner, D.~Malyshev, O.~Ruchayskiy, T.~Slatyer and N.~Weiner for their help in resolving this issue.} 
\begin{equation}
{\mathcal F}(E_{\gamma},\theta) \propto \exp\bigg[-\frac{\theta^2}{2 \, \sigma^2(E_{\gamma})}\bigg],
\label{psf}
\end{equation}
where $\theta$ is the difference between the measured and actual directions of the observed gamma ray. We select values for $\sigma(E_{\gamma})$ which reproduce the 68\% and 95\% containment radai, as quoted by the Fermi Collaboration~\cite{psf}. The angle within which 68\% of the gamma rays are reconstructed is well fit (in degrees) by
\begin{eqnarray}
\log_{10} [P_{68}(E_{\gamma})] &=& -0.276 - 0.674 \log_{10}[E_{\gamma}/1\,{\rm GeV}], \,\,\,\,\,\,\,\,\, E_{\gamma} < 20 \, {\rm GeV} \nonumber \\
\log_{10} [P_{68}(E_{\gamma})] &=& -0.785 - 0.283 \log_{10}[E_{\gamma}/1\,{\rm GeV}], \,\,\,\,\,\,\,\,\, E_{\gamma} > 20 \, {\rm GeV}.
\end{eqnarray}
We also include a small correction to account for the non-gaussianity of the LAT's PSF. This has little effect on our results, however.


At angular distances of more than 1-2$^{\circ}$ from the Galactic Center, the disk component of the emission can be clearly and easily separated from the other components. For this reason, we begin by studying only the region of the sky with $\sqrt{l^2 + b^2} > 2^{\circ}$. For a given angular annulus, we fit the data to a combination of the three components described above, and consider one energy bin at a time (distributed logarithmically in energy, with ten bins per decade). For each combination of energy and distance from the Galactic Center, we find the combination of the following parameters that yields the best overall chi-square fit:  the intensity of the disk emission for $l < 0^{\circ}$, the intensity of the disk emission for $l > 0^{\circ}$, the width of the disk emission for $l < 0^{\circ}$, the width of the disk emission for $l > 0^{\circ}$, and the intensity of the spherically symmetric emission. For simplicity, we fix the intensity of each point source assuming that it has a power-law spectrum with an index and intensity as described in the Fermi First Source Catalog (the central values that are quoted). At times, this simplifying choice will not be supported by the data (departures from power-law behavior are evident for some of the point sources), but this impacts our overall results only slightly. 

\begin{figure}[!htb]
\resizebox{15.0cm}{!}{\includegraphics{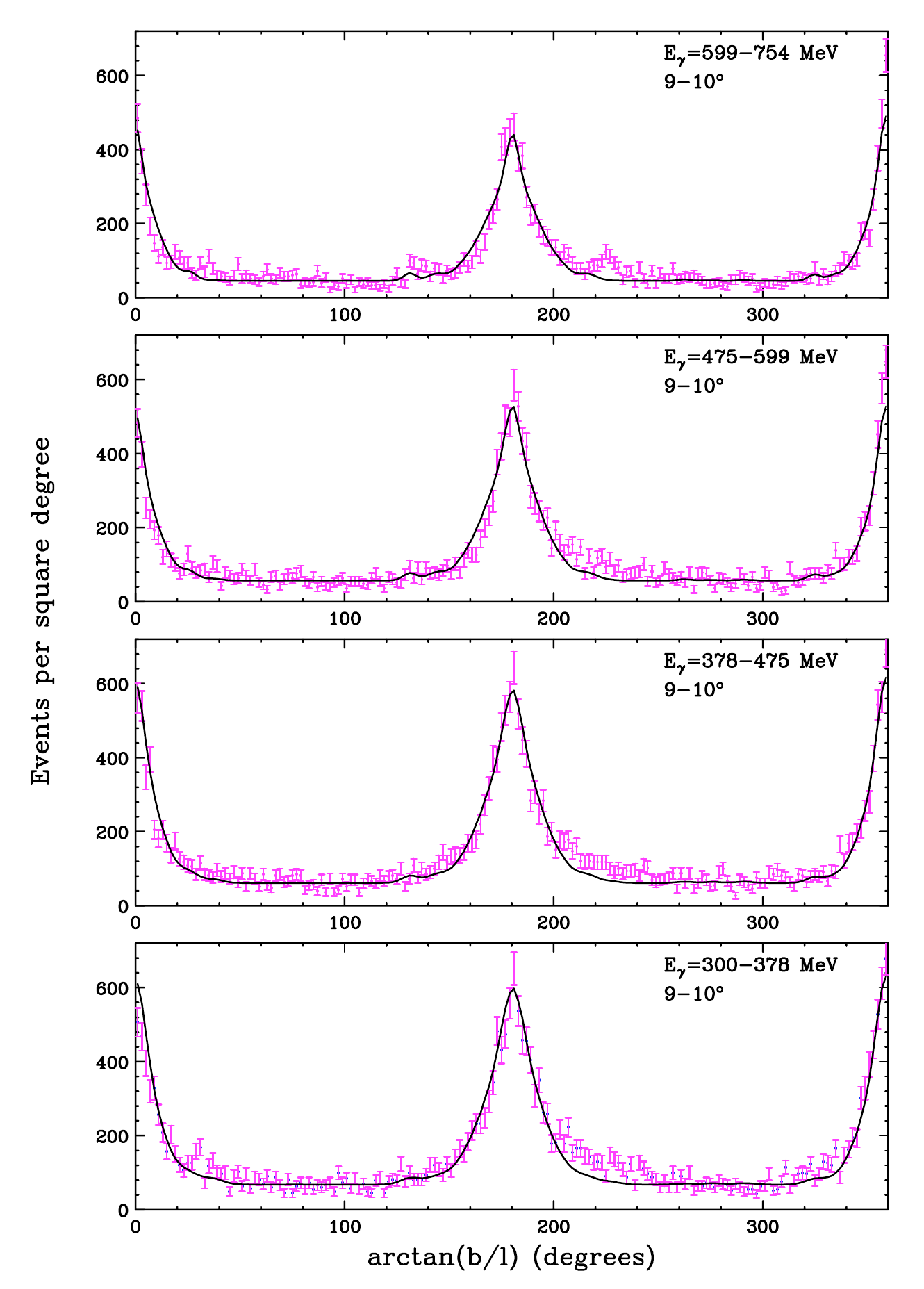}} \\
\caption{The best fit model for the emission in the four lowest energy bins from the region between 9 and 10 degrees from the Galactic Center (the region between the two circles shown in Fig.~\ref{ptsources}), compared to the observations of the FGST. See text for details.}
\label{angular1}
\end{figure}

\begin{figure}[!htb]
\resizebox{15.0cm}{!}{\includegraphics{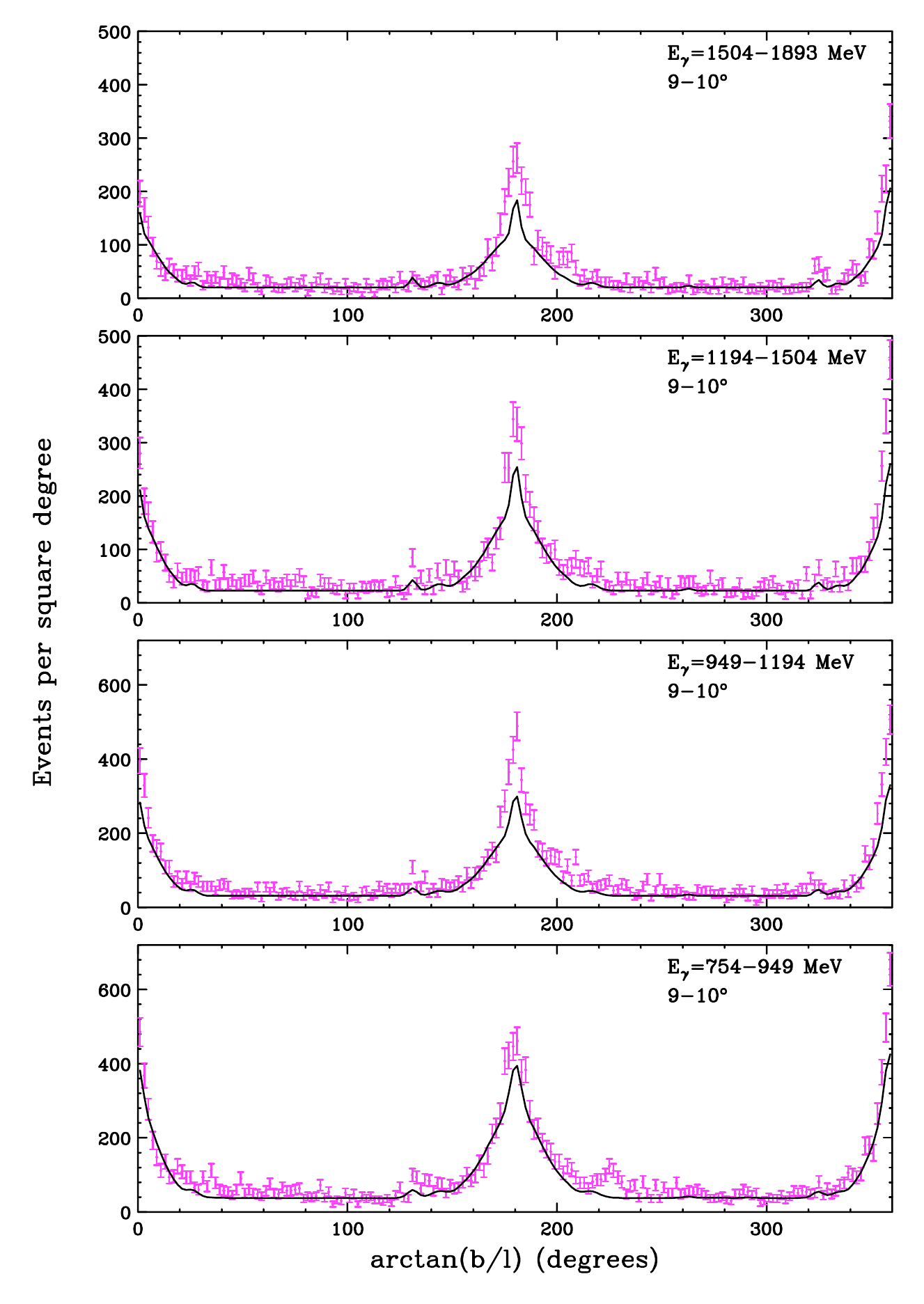}} \\
\caption{The same as Fig.~\ref{angular1}, but for higher energy bins, as labeled.}
\label{angular2}
\end{figure}

\begin{figure}[!htb]
\resizebox{15.0cm}{!}{\includegraphics{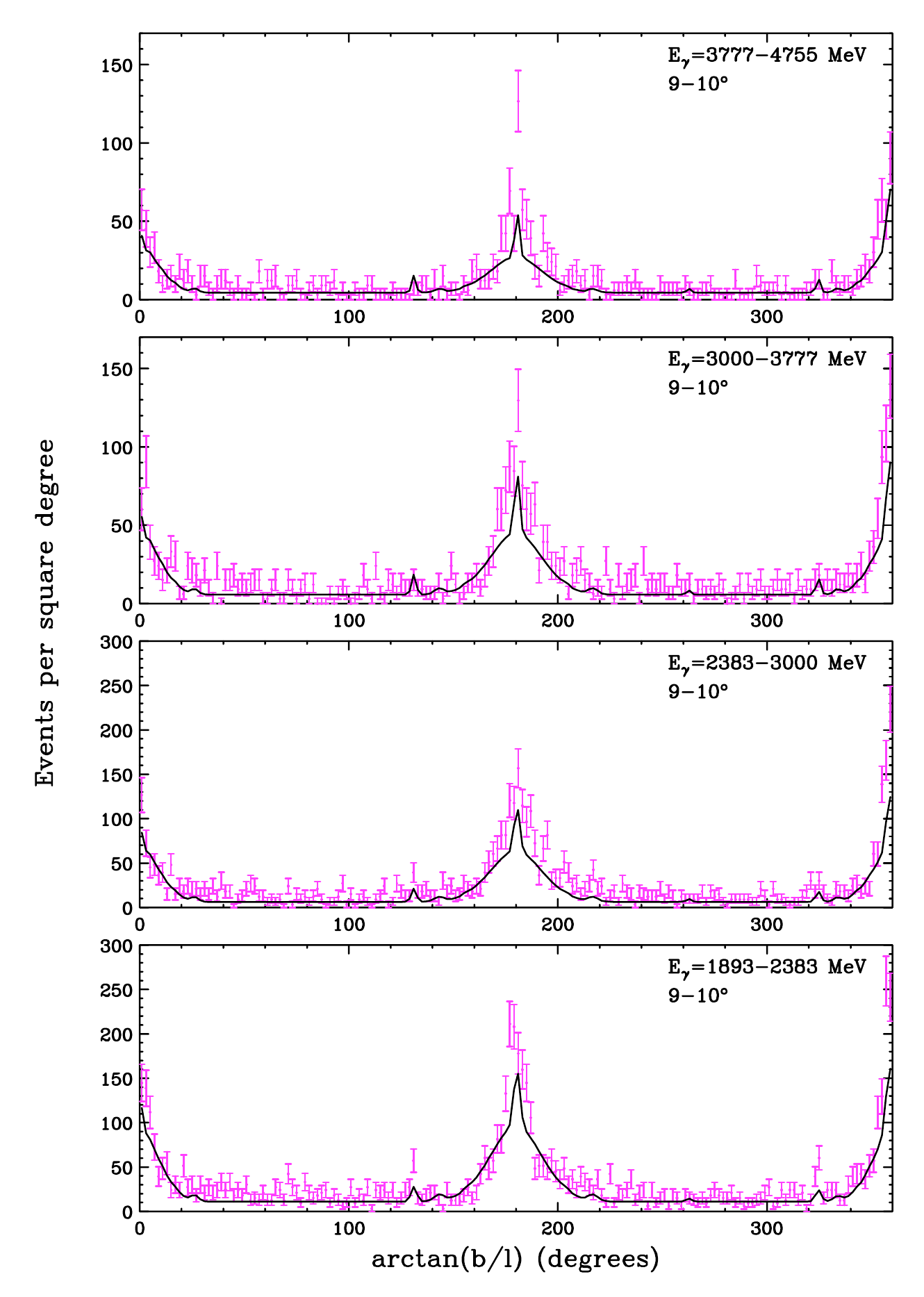}} \\
\caption{The same as Fig.~\ref{angular1}, but for higher energy bins, as labeled.}
\label{angular3}
\end{figure}

\begin{figure}[!htb]
\resizebox{15.0cm}{!}{\includegraphics{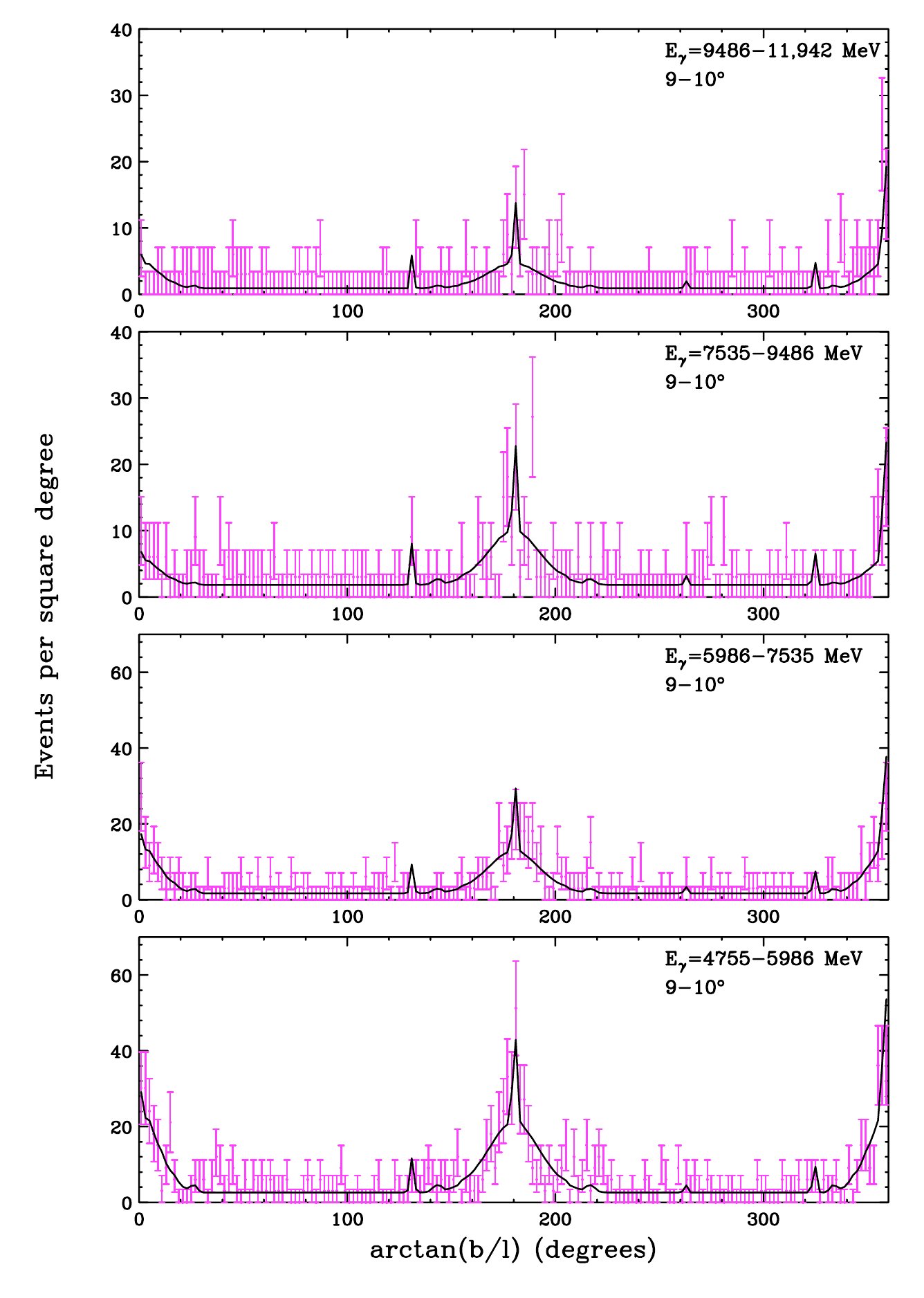}} \\
\caption{The same as Fig.~\ref{angular1}, but for higher energy bins, as labeled.}
\label{angular4}
\end{figure}

In Figs.~\ref{angular1}-\ref{angular4}, we show the angular distribution (meaning with the direction from the Galactic Center) of the (front converting) events observed by FGST between 9 and 10 degrees away from the Galactic Center in the first 16 energy bins (300 MeV to 11.94 GeV), and compare this to that predicted for our best-fit model parameters. In each frame, the emission from the disk is clearly apparent and in some cases emission from individual point sources can be seen (compare to the point sources between the two circles shown in Fig.~\ref{ptsources}). The data shown corresponds to that collected between August 4, 2008 and August 12, 2010.

The most significant discrepancies between our best fit model and the data, as shown in Figs.~\ref{angular1}-\ref{angular4} (and also over other angular ranges), results from our simplistic treatment of point sources. Because we have not allowed the intensity of the point source contributions to float in this fit, but rather have fixed them to the overall flux and power-law spectral index listed in the FGST First Source Catalog, there are evident examples in which either brighter or dimmer point source emission would provide a better fit, relative to that included in our simple model. Notice, for example, the $E_{\gamma}=$599-754 MeV and 754-949 MeV frames of Figs.~\ref{angular2}-\ref{angular3}, at approximately $\arctan(b/l) \approx 220-230^{\circ}$. Here the spectrum of the responsible nearby point source (clearly identifiable in Fig.~\ref{ptsources} as 1FGL J1802.5-3939, located at $l=-7.5581^{\circ}$, $b=-8.3935^{\circ}$) evidently exceeds that predicted by the best fit power-law. If it were not for considerations of computation time, we could fit for the individual intensities of each point source in each energy bin. Given that less than 7\% of the total gamma ray flux in the inner $\pm 15^\circ$ window is associated with resolved point sources, this choices does not significantly impact our extraction of the spectral shapes or intensities of the gamma ray emission from the disk or from the spherically symmetric components.

\begin{figure}[!htb]
\begin{center}
\resizebox{18.4cm}{!}{\includegraphics{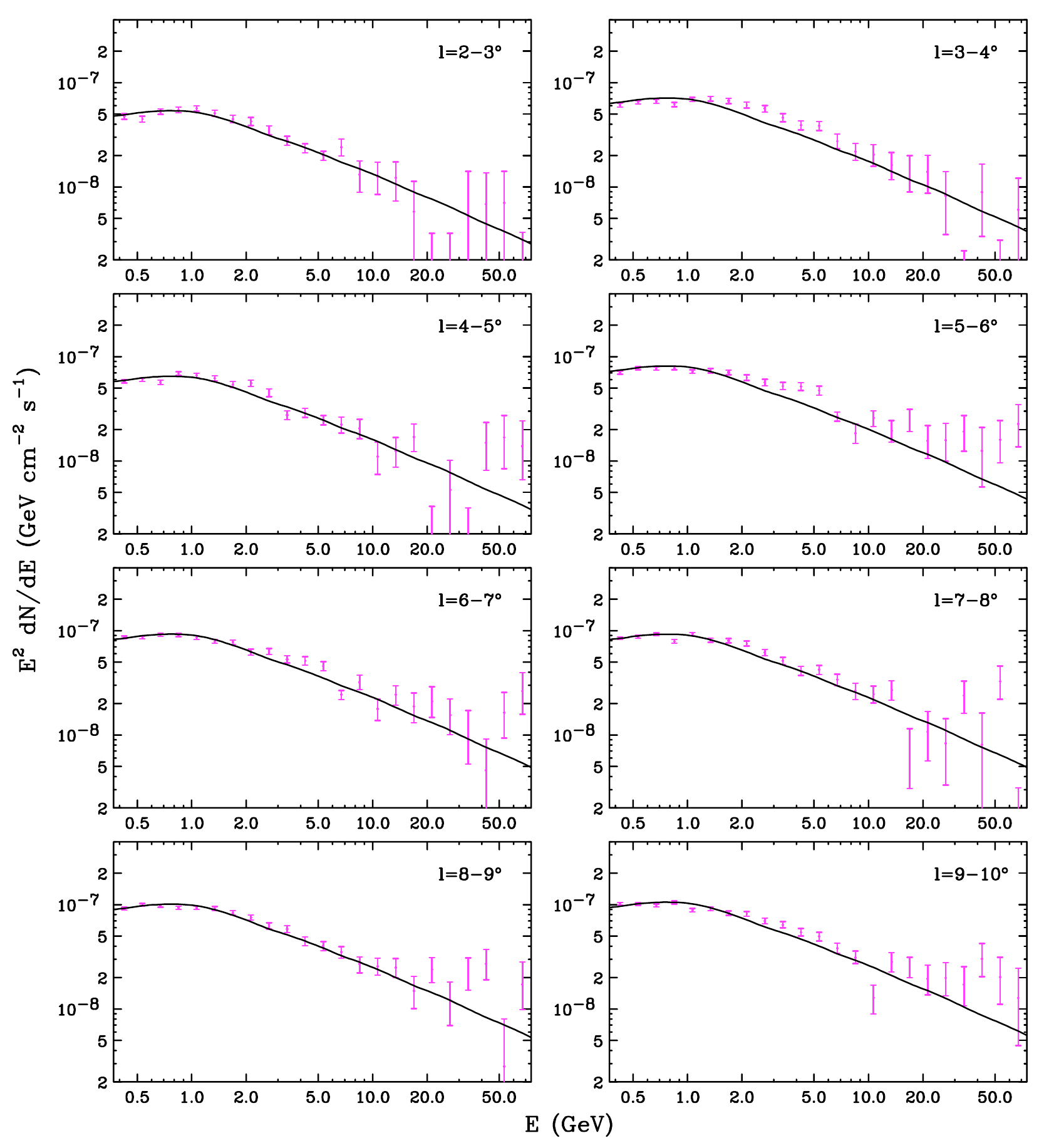}} \\
\caption{The spectrum of emission from the disk for $l>2^{\circ}$. Note that the spectral shape and intensity varies only slightly along the plane. The solid line is the shape predicted from the combination of pion decay and inverse Compton scattering (a small contribution from Bremsstrahlung may also be present), as shown in Fig.~\ref{galprop}. The overall normalization of this emission varies by less than a factor of 2 among the frames. See text for details.}
\label{left}
\end{center}
\end{figure}

\begin{figure}[!htb]
\begin{center}
\resizebox{18.4cm}{!}{\includegraphics{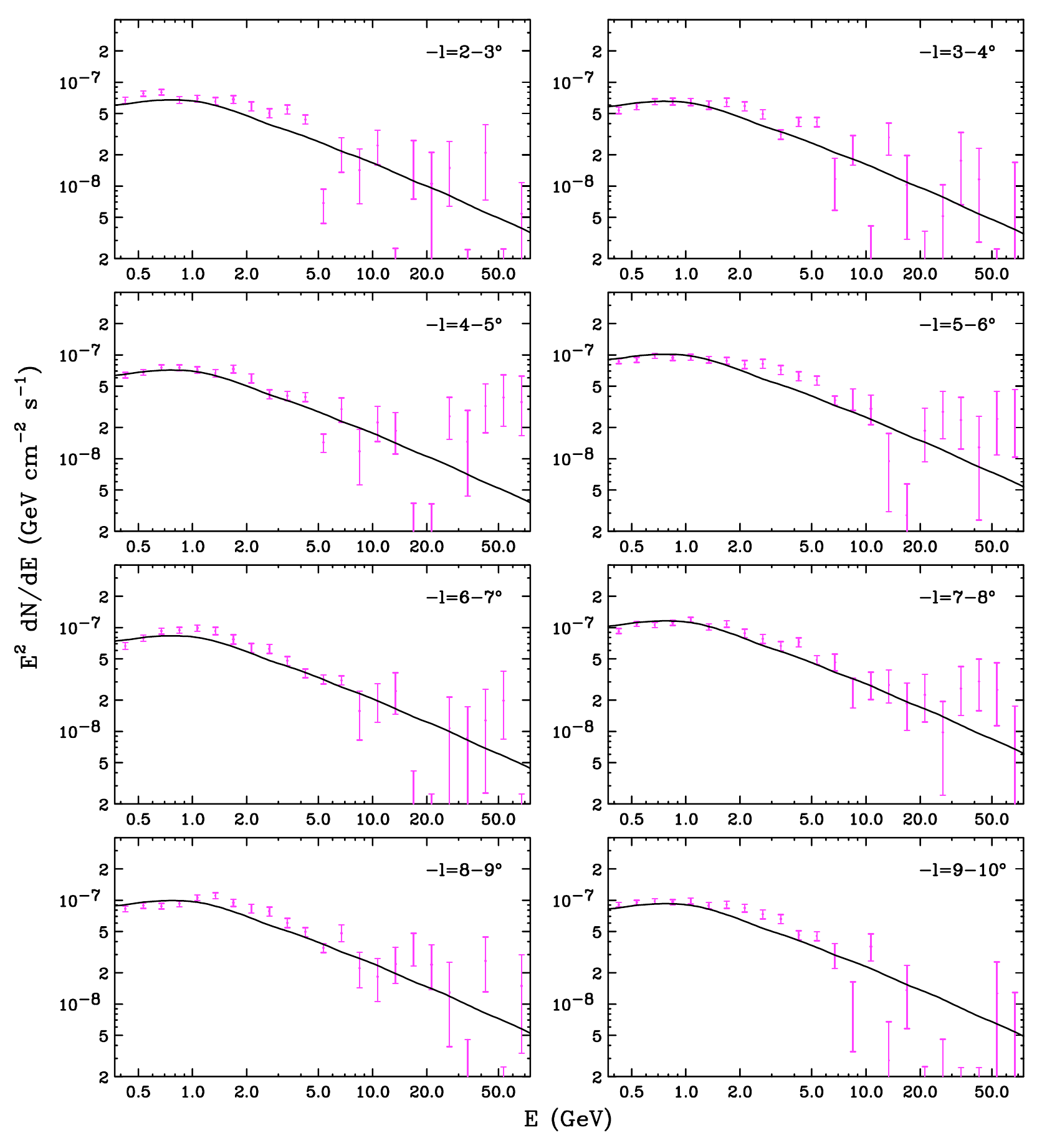}} \\
\caption{The spectrum of emission from the disk for $l<-2^{\circ}$. Note that the spectral shape and intensity varies only slightly along the plane. The solid line is the shape predicted from the combination of pion decay and inverse Compton scattering (a small contribution from Bremsstrahlung may also be present), as shown in Fig.~\ref{galprop}. The overall normalization of this emission varies by less than a factor of 2 among the frames. See text for details.}
\label{right}
\end{center}
\end{figure}

\begin{figure}[!htb]
\begin{center}
\resizebox{18.4cm}{!}{\includegraphics{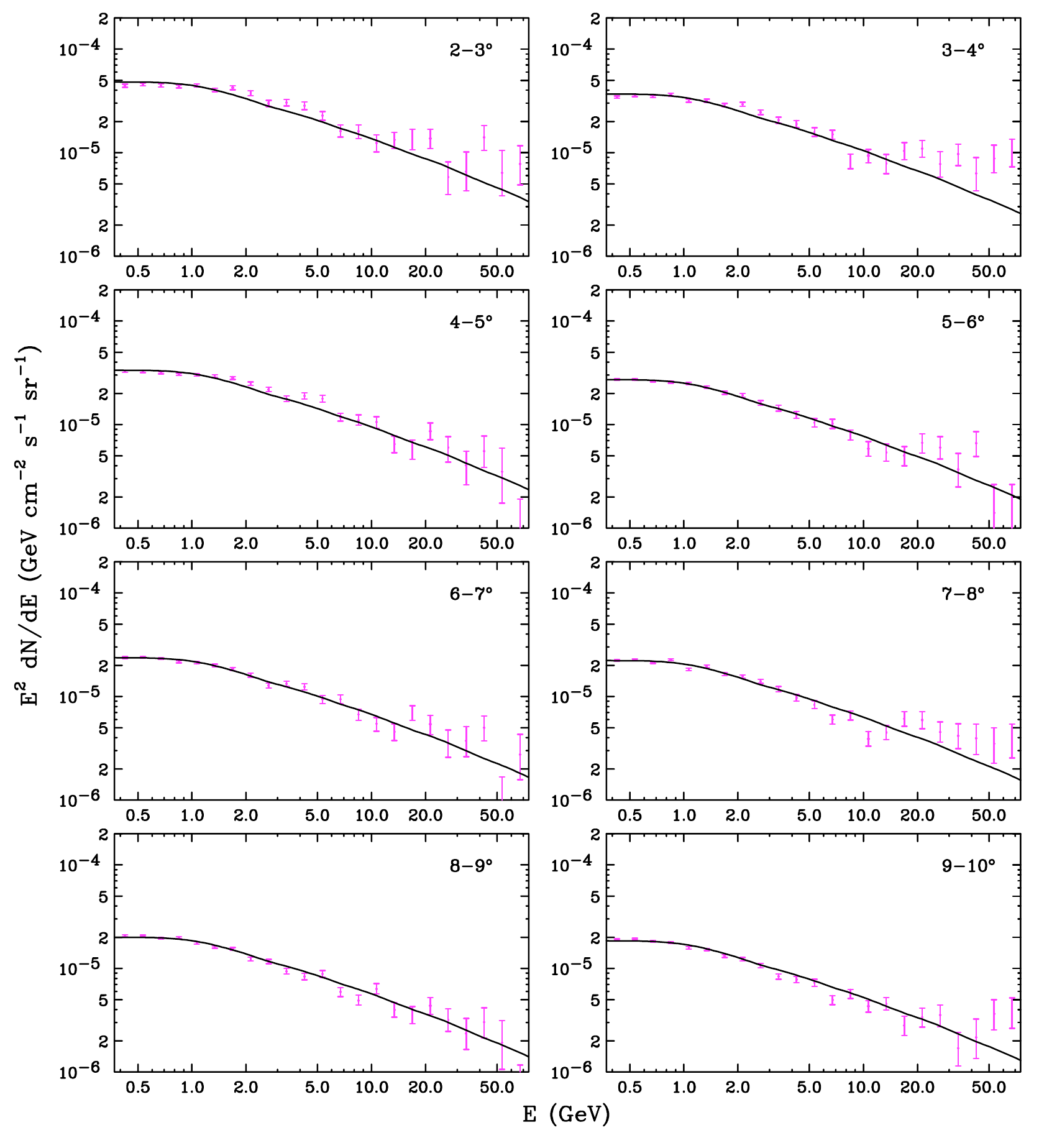}} \\
\caption{The spectrum of the emission that is distributed with spherical symmetry around the Galactic Center (not disk-like or associated with known point sources), for different ranges of angles from the Galactic Center (each frame shows the emission from a given angular annulus).  Although the spectral shape does not discernibly change with distance to the Galactic Center, the overall intensity does vary, becoming brighter closer to the Inner Galaxy. The solid line is the shape predicted from the combination of pion decay and inverse Compton scattering (a small contribution from Bremsstrahlung could also be present), as shown in Fig.~\ref{galprop}, but with a larger relative flux of inverse Compton gamma rays than in Figs.~\ref{left}-\ref{right} (the spectral shape of the bulge emission contains somewhat more high energy emission than that from the disk).}
\label{flat}
\end{center}
\end{figure}

We have repeated this procedure to derive the spectra of the disk and spherically symmetric components of the gamma ray emission over various regions of the Inner Galaxy (between 2$^\circ$ and 10$^{\circ}$ from the Galactic Center). The results are shown in Figs.~\ref{left}-\ref{flat}. We find that the emission from the disk has a fairly uniform spectral shape and overall intensity, varying only slightly along the disk (with $l$). The spherically symmetric emission also shows no discernible variation in the spectral shape, but does become steadily brighter as we move closer to the Galactic Center.

\begin{figure}[!htb]
\resizebox{14.0cm}{!}{\includegraphics{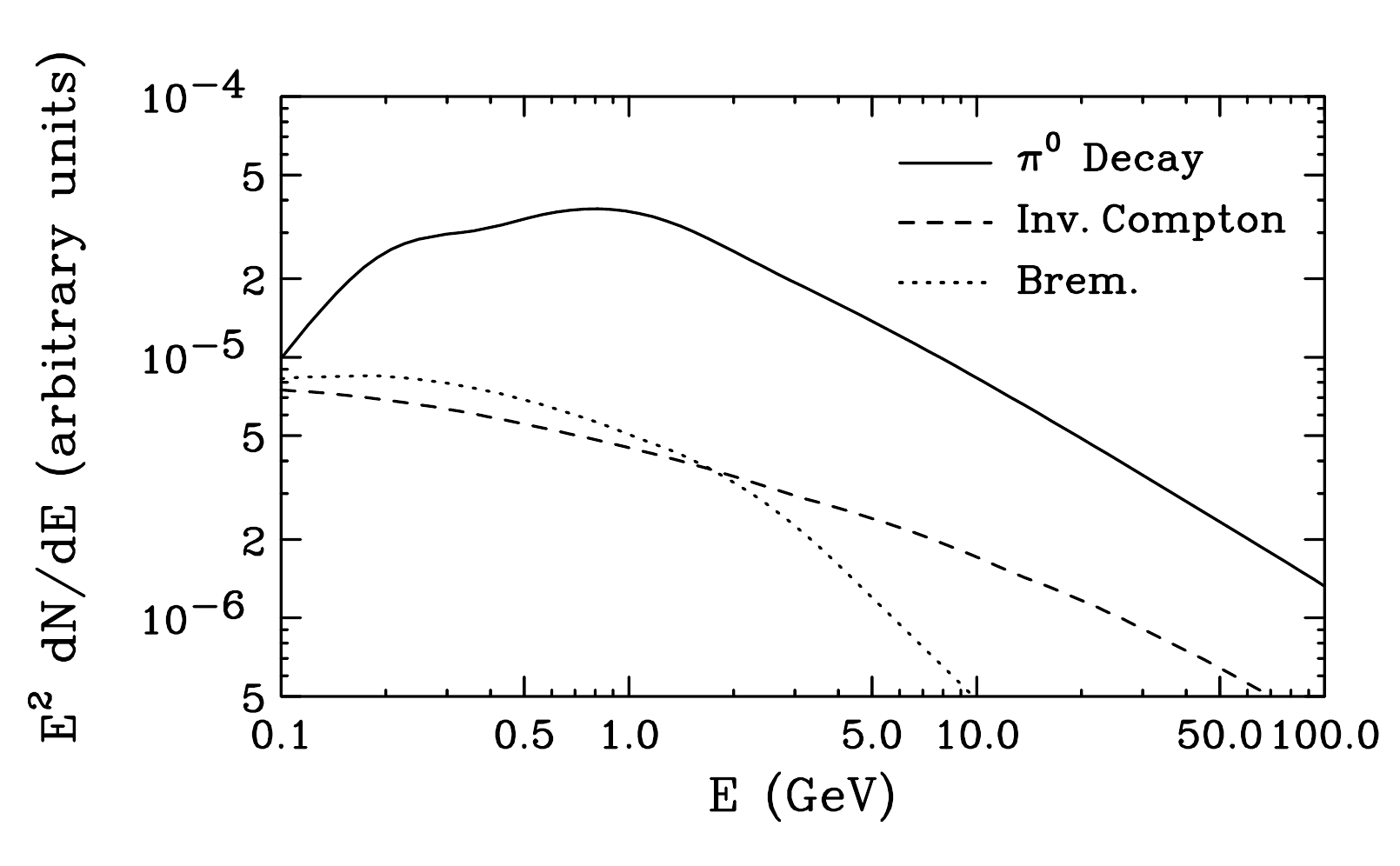}} \\
\caption{The predicted spectral shapes of gamma rays from pion decay, inverse Compton scattering, and Bremsstrahlung in the region around the Galactic Center, as generated using the publicly available code GALPROP~\cite{galprop}.}
\label{galprop}
\end{figure}

The spectra shown in Figs.~\ref{left}-\ref{flat} can be easily accounted for with known emission mechanisms. In particular, their spectral shapes are consistent with being dominated by gamma rays from neutral pion decay, with a smaller but not insignificant contribution from inverse Compton scattering (a small contribution from Bremsstrahlung may also be present). In Fig.~\ref{galprop}, we show the shapes of the spectra predicted from these emission mechanisms, as calculated using the publicly available code GALPROP~\cite{galprop}. The solid lines shown in Figs.~\ref{left}-\ref{right} correspond to the predicted spectrum from pion decay and inverse Compton scattering, with relative normalizations as shown in Fig.~\ref{galprop}. This provides a good overall fit to the observed spectra from the disk. In Fig.~\ref{flat}, the solid line again represents the contribution from pion decay and inverse Compton scattering, but with a larger fraction of the emission from inverse Compton scattering (7.15 times larger, relative to the pion component, than in the disk). Again, this provides a good overall fit to the observed spectra. The greater relative contribution from inverse Compton scattering (or equivalently, the lesser relative contribution from pion decay) is likely the consequence of lower gas densities away from the Galactic Disk.

The variation of the intensity of the observed disk emission with distance from the Galactic Center is well described by a density profile of emission which scales with $r^{-1.55}$ within the inner 2 kiloparsecs or so, where $r$ is the distance to the Galactic Center (see Fig.~\ref{profile}). This component is likely associated primarily with gas in the Galactic Bulge. Given the agreement of the observed spectrum with that predicted from pion decay and inverse Compton scattering, we see no evidence of any exotic component among the bulge emission (outside of 2$^{\circ}$ from the Galactic Center).


\newpage

\begin{figure}[!htb]
\resizebox{14.0cm}{!}{\includegraphics{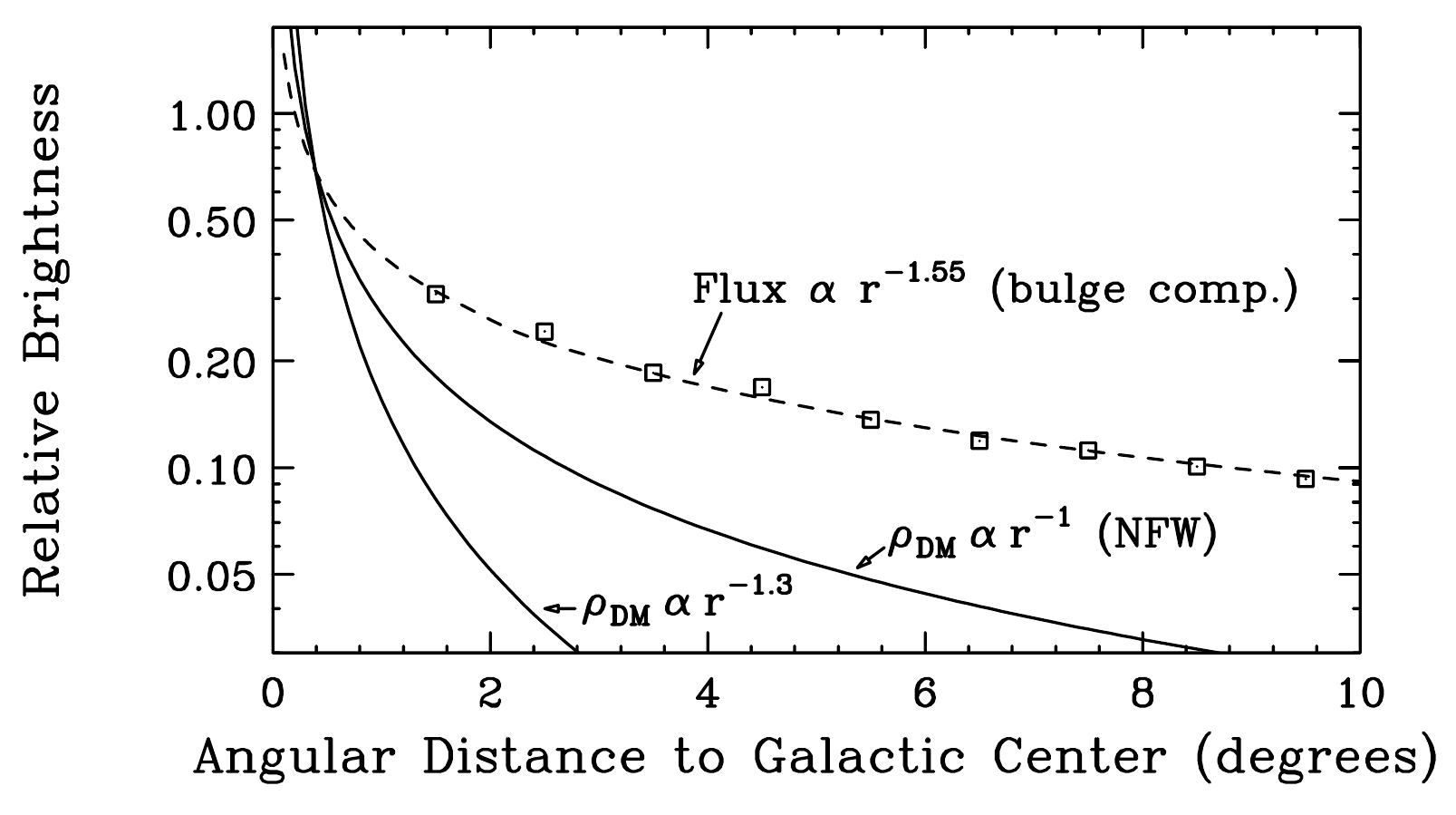}} \\
\caption{The relative brightness, integrated along the line-of-sight, of the emission that is distributed with spherical symmetry around the Galactic Center ({\it ie.}~the bulge component), as a function of the distance to the Galactic Center (squares), compared to that predicted for emission that is distributed as $r^{-1.55}$, where $r$ is the distance to the Galactic Center (dashed line). For comparison, we also show the distribution for emission from dark matter annihilations using a NFW ($\gamma=1$) halo profile or a NFW-like profile with $\gamma=1.3$ (solid).}
\label{profile}
\end{figure}

\section{The Inner Two Degrees Around The Galactic Center}
\label{inner}

\begin{figure}[!htb]
\resizebox{18.4cm}{!}{\includegraphics{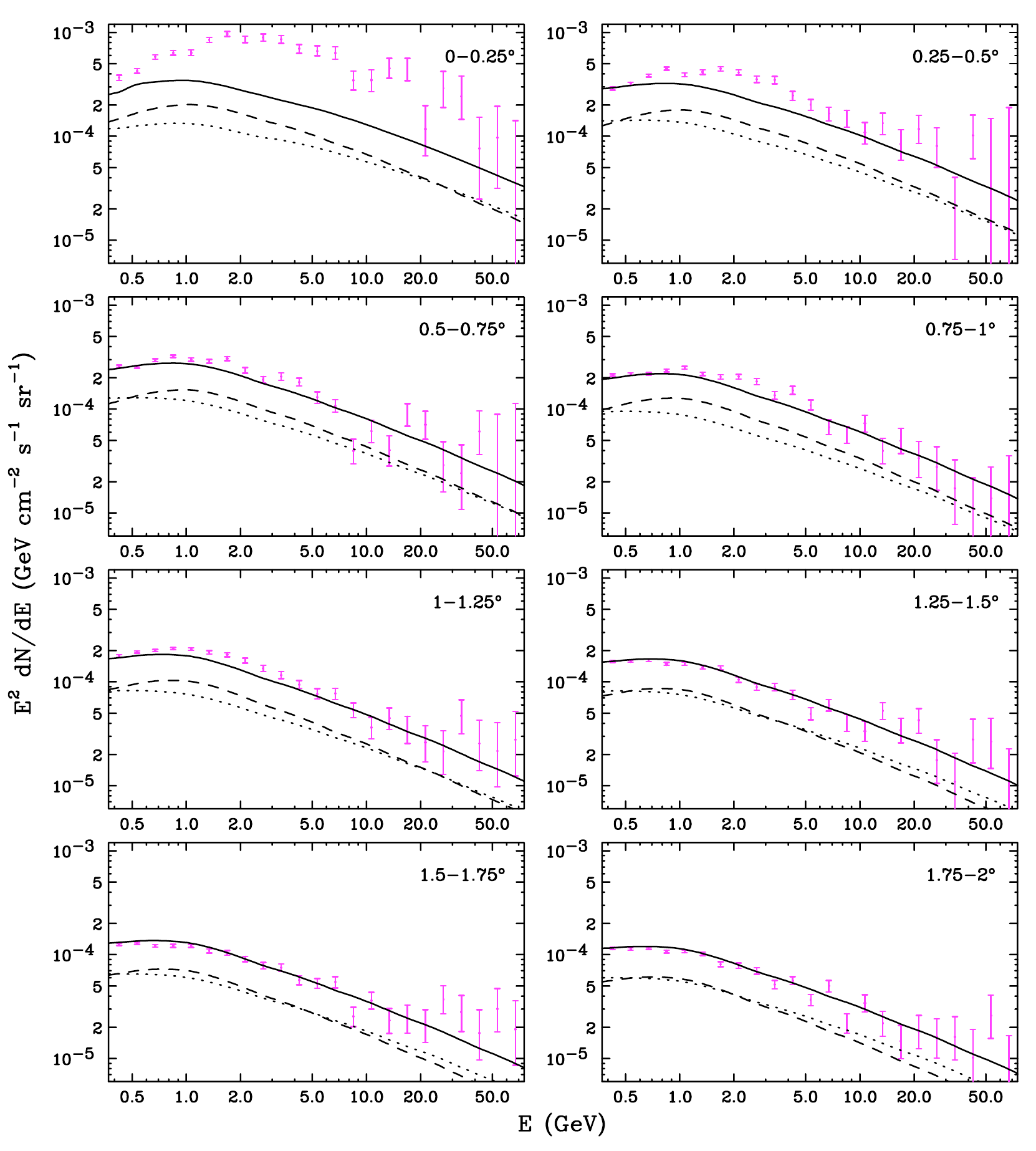}} \\
\caption{The total observed gamma ray spectrum in various ranges of the angular distance from the Galactic Center, compared to the bulge (dotted) and disk (dashed) components of our background model (the solid line denotes the sum of the bulge and disk components). Outside of 1.25$^{\circ}$ from the Galactic Center, this model describes the data very well. Closer to the Galactic Center, however, the spectral shape of the observed emission is significantly different, peaking at 1-5 GeV. See text for further details.}
\label{total}
\end{figure}

Within 1-2$^{\circ}$ of the Galactic Center, it is more difficult to clearly separate the spherically symmetric contributions from those originating from the disk. Instead, we compare in Fig.~\ref{total} the total emission in the innermost angular regions to a model consisting of the extrapolated disk emission (which was found to be relatively constant between $|l|=2-10^{\circ}$) and bulge emission increasing with angle according to the best fit $r^{-1.55}$ profile. In each frame, the dotted and dashed lines represent the bulge and disk components of our model, respectively, while the solid line denotes the sum of these components. For regions more than approximately $1.25^{\circ}$ away from the Galactic Center, this model describes the total observed spectra very well. Inside of this radius, however, the spectral shape changes considerably, becoming considerably more intense over the range of about 1-5 GeV. No combination of the spectra from the mechanisms shown in Fig.~\ref{galprop} can account for this transition.

\begin{figure}[!htb]
\resizebox{19.0cm}{!}{\includegraphics{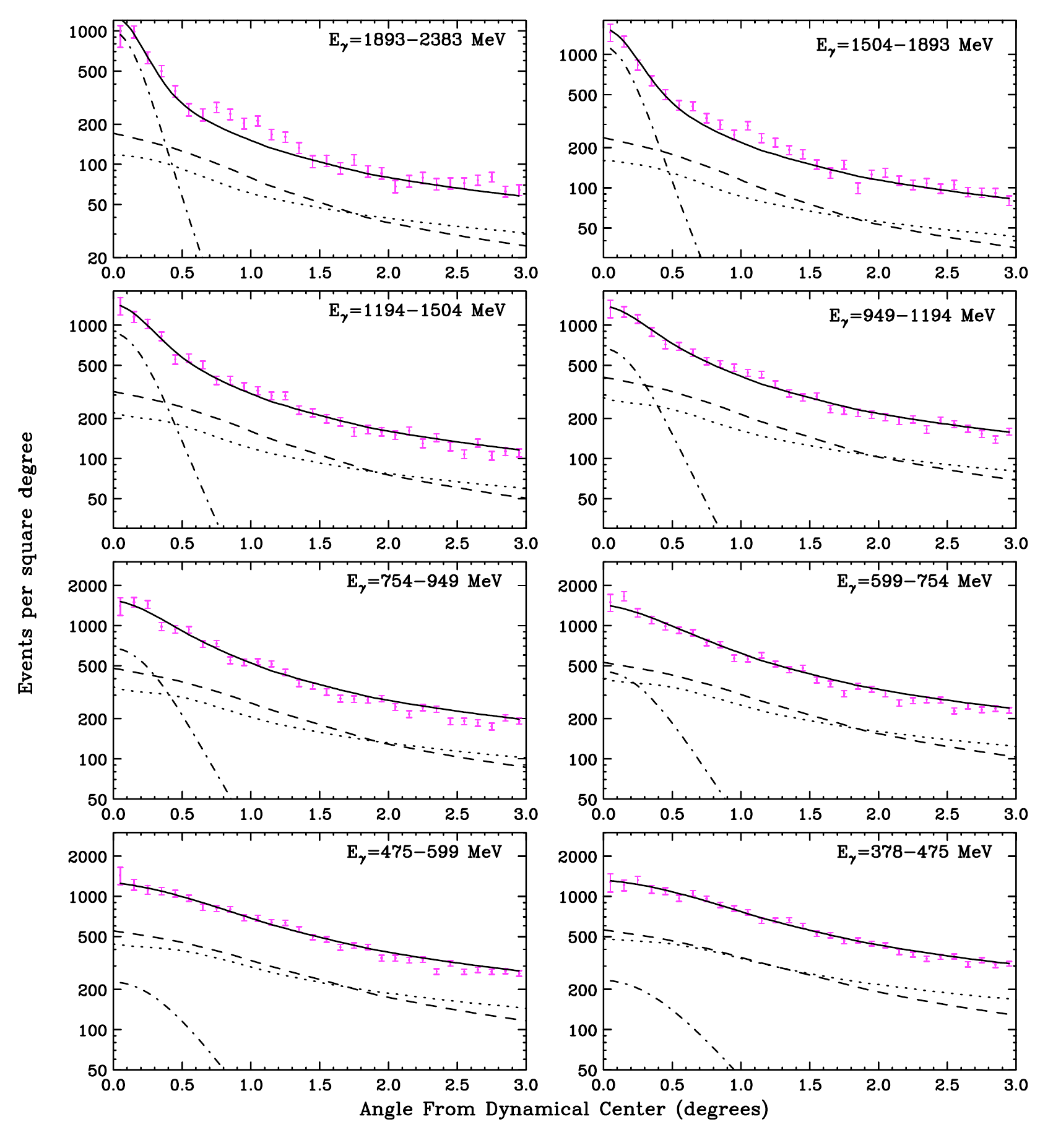}} \\
\caption{The angular profile of the emission that is distributed with spherical symmetry around the Galactic Center. The bulge, disk, central point source, and bulge+disk+central point source components are shown as dotted, dashed, dot-dashed, and solid lines, respectively.}
\label{pro1}
\end{figure}

\begin{figure}[!htb]
\resizebox{19.4cm}{!}{\includegraphics{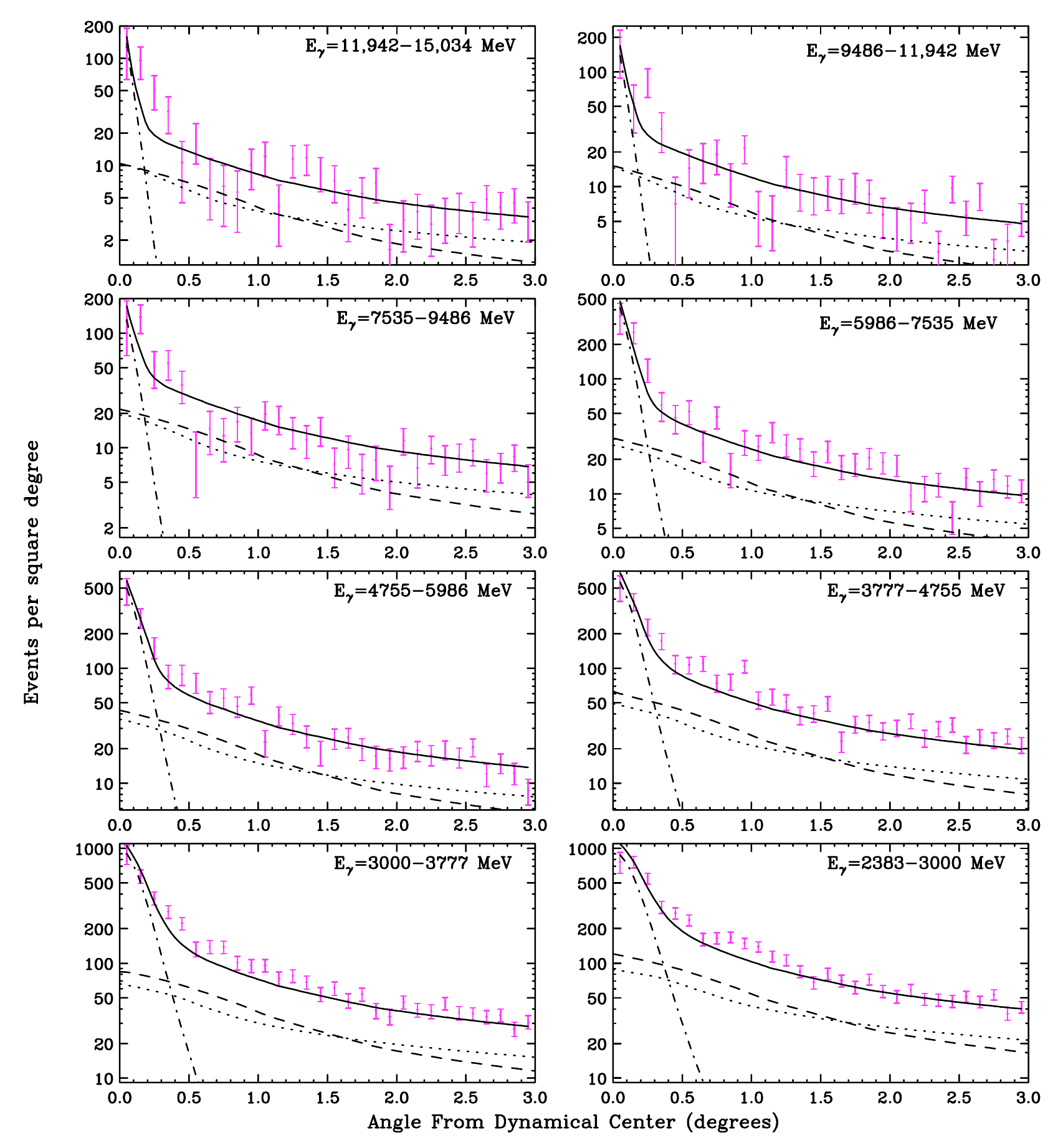}} \\
\caption{The angular profile of the emission that is distributed with spherical symmetry around the Galactic Center. The bulge, disk, central point source, and bulge+disk+central point source components are shown as dotted, dashed, dot-dashed, and solid lines, respectively.}
\label{pro2}
\end{figure}

To further elucidate this observed transition in the inner $\sim$1.25$^{\circ}$ of our Galaxy, we plot in Figs.~\ref{pro1}-\ref{pro2} the angular profile of the observed emission in 16 energy bins. Again, the dotted and dashed lines denote the bulge and disk components respectively. In addition, we show as a dot-dashed line the best-fit contribution from a point source located at the dynamical center of the Galaxy (Sgr A$^{\star}$, or emission associated with the supermassive black hole). In each frame, the solid line represents the sum of bulge, disk, and point source contributions.

\begin{figure}[!htb]
\resizebox{8.85cm}{!}{\includegraphics{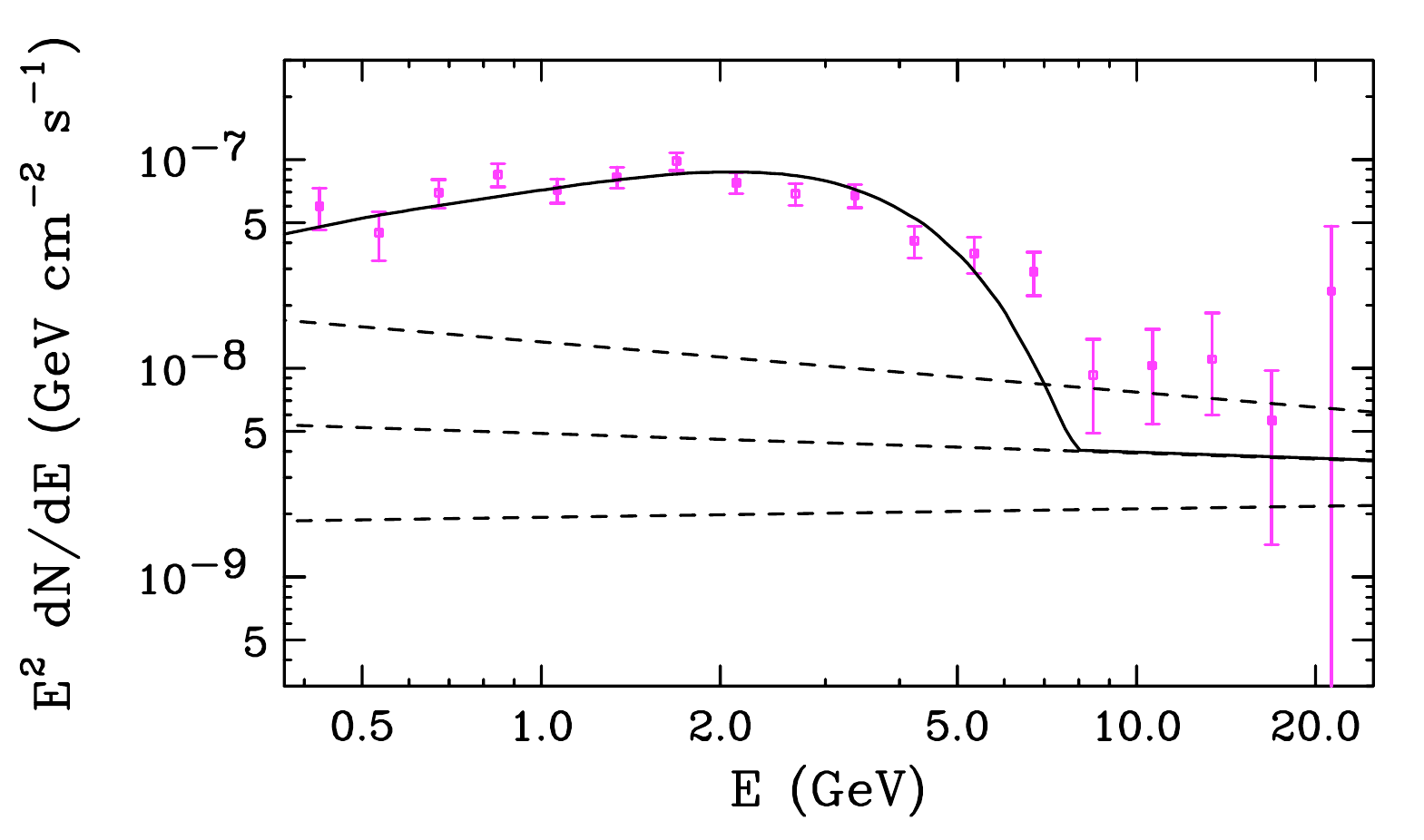}} 
\resizebox{8.85cm}{!}{\includegraphics{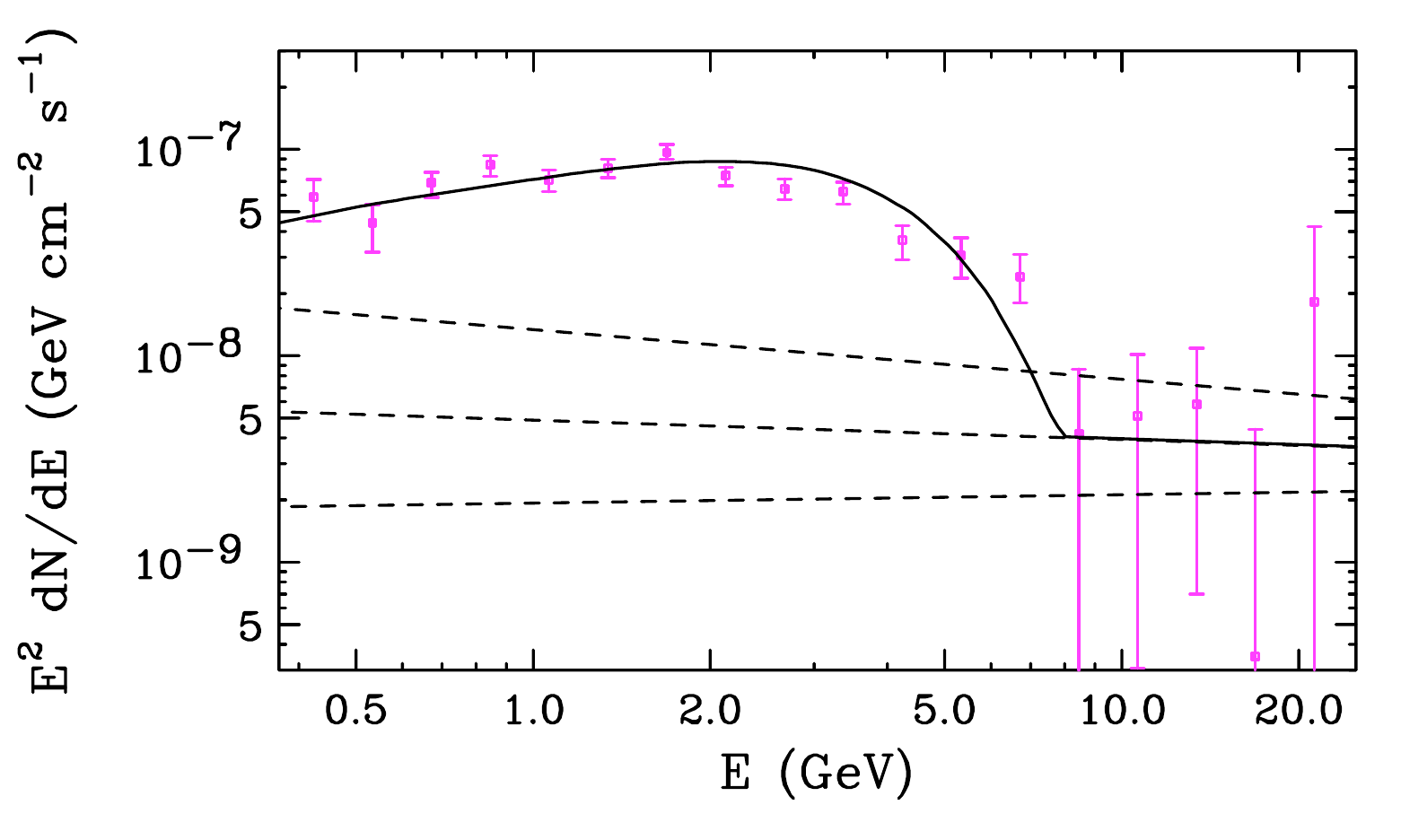}}
\caption{The spectrum of point-like emission from the Galactic Center (as calculated using disk, bulge and point source templates, but without any spatially extended contribution from dark matter). The spectrum shown has been corrected to account for the finite point spread function of FGST. Shown as dashed lines is the power-law extrapolation of the Galactic Center point source as measured by HESS, $dN_{\gamma}/dE_{\gamma}=[2.55 \pm 0.46] \times 10^{-12} \,{\rm TeV}^{-1} {\rm cm}^{-2} {\rm s}^{-1} \times [E_{\gamma}/1\,{\rm TeV}]^{-2.1 \pm 0.14}$. In the right frame, the spectrum has been corrected to account for emission from the Galactic Ridge, as measured by HESS. See text and Fig.~\ref{ridge} for details.}
\label{spectracen}
\end{figure}

For each energy bin, we do a fit to the angular profile (as shown in Fig.~\ref{pro1}-\ref{pro2}), for a combination of disk and bulge components as shown, plus each point source in the Fermi First Source Catalog, including the central source. In calculating the quality of the fits, we include a small 3\% systematic error, which is intended to account for the spatial variation in the disk and bulge emission from, for example, unresolved point sources. The best fit we find using these templates provides a fairly reasonable description of the data ($\chi^2 \approx 1.1$ per degree of freedom). In the left frame of Fig.~\ref{spectracen}, we show the spectrum of the point-like emission from the Galactic Center obtained with this fit. Shown for comparison as dashed lines is the spectrum from this source extrapolated from higher energy measurements made by the ground-based gamma ray telescope HESS~\cite{hess} and other ground based gamma ray telescopes~\cite{otheract}. Over the energy range of approximately 200 GeV to 10 TeV, HESS's measurements of Sgr A$^{\star}$ reveal a spectrum given by $dN_{\gamma}/dE_{\gamma}=[2.55 \pm 0.46] \times 10^{-12} \,{\rm TeV}^{-1} {\rm cm}^{-2} {\rm s}^{-1} \times [E_{\gamma}/1\,{\rm TeV}]^{-2.1 \pm 0.14}$ (the middle dashed line denotes the central value for the HESS measurement, while the upper and lower lines represent the 1-$\sigma$ upper and lower values for both the normalization and spectral slope).

\begin{figure}[!htb]
\resizebox{10.0cm}{!}{\includegraphics{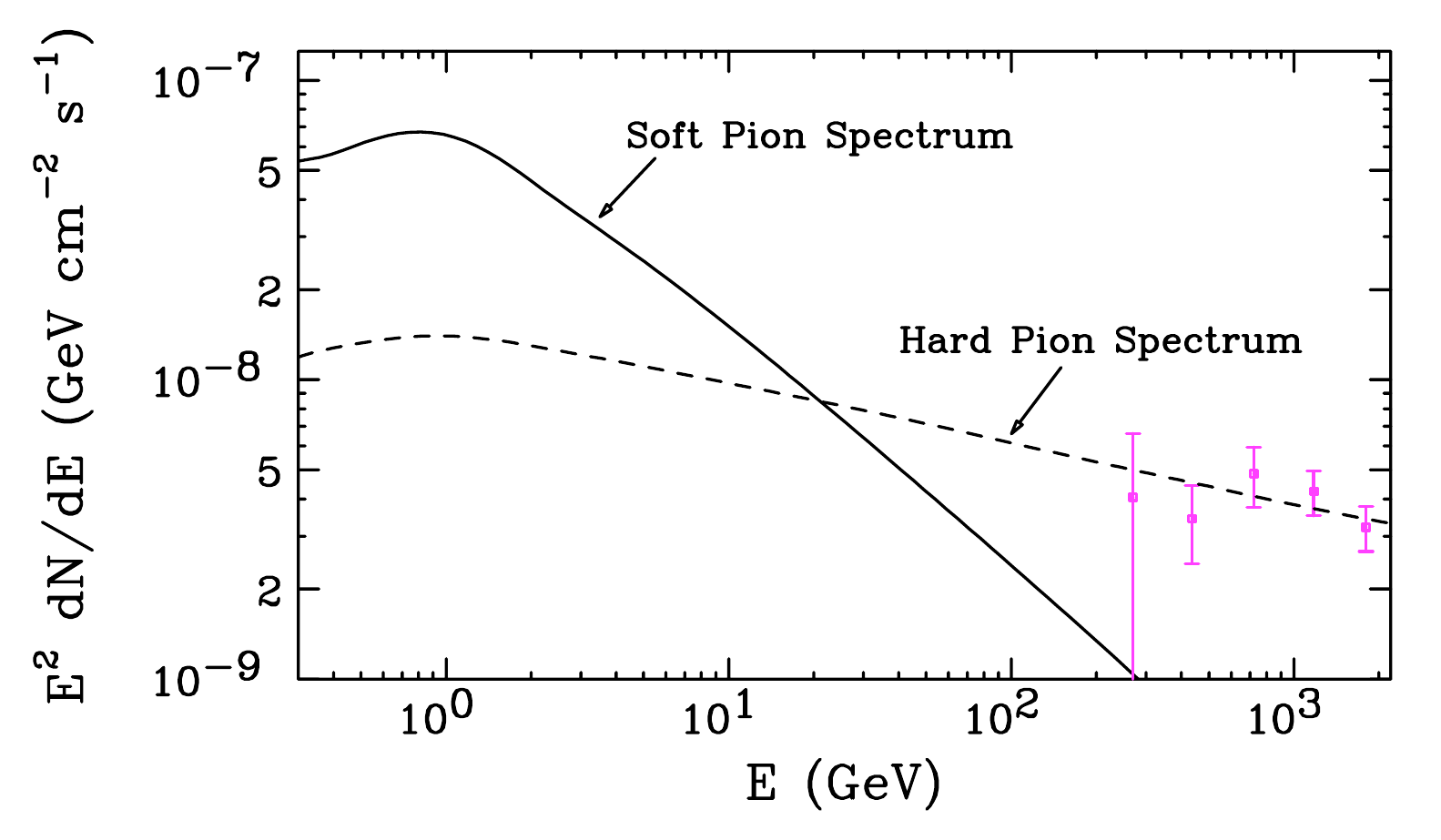}} \\
\caption{The spectral shape of gamma rays from pion decay as observed throughout most of the inner 10$^{\circ}$ around the Galactic Center (soft pion spectrum) and from pion decay with a proton spectrum selected to reproduce the Galactic Ridge emission observed by HESS (hard pion spectrum). Also shown are the several lowest energy error bars as measured by HESS.}
\label{ridge}
\end{figure}

Observations from HESS have also identified extended emission associated with the inner Galactic Ridge~\cite{ridge}, with a spectrum given by $dN_{\gamma}/dE_{\gamma}=[1.73 \pm 0.48] \times 10^{-8} \,{\rm TeV}^{-1} {\rm cm}^{-2} {\rm s}^{-1} {\rm sr}^{-1} \times [E_{\gamma}/1\,{\rm TeV}]^{-2.29 \pm 0.09}$, over the region of $|l| < 0.8^{\circ}$ and $|b| < 0.3^{\circ}$. This is generally interpreted as the product of hadronic cosmic rays producing pions through collisions with molecular clouds. If so, it implies that the cosmic ray spectrum in the Inner Galaxy includes a component with a spectral component that is harder than is found elsewhere. In particular, we find that the observed gamma ray spectrum of the ridge requires the responsible cosmic rays protons to possess a spectral index approximately given by $dN_{p}/dE_{p} \propto E_{p}^{-1.89}$. In Fig.~\ref{ridge}, we compare the gamma ray spectrum from protons with this spectral index to that found elsewhere within the inner 10$^{\circ}$ around the Galactic Center. To account for this harder emission, we substitute a fraction (chosen to maintain the overall normalization at low energies) of the pion emission in the bulge component with that with the spectral shape required to produce the ridge emission observed by HESS. The effect of this correction is almost entirely negligible below $\sim$5-6 GeV, but has a noticable impact at higher energies. In the right frame of Fig.~\ref{spectracen}, we show the spectrum of point-like emission after correcting for the ridge emission. Once this is acconted for, we find that the spectrum of the central point source is in good agreement with the extrapolation of HESS's power-law spectrum at energies above about 8 GeV. At lower energies, however, the observed spectrum is brighter by up to roughly an order of magnitude, compared to the power-law extrapolation. While we cannot rule out the possibility that the spectrum of this source strongly departs from the extrapolation below $\sim$8 GeV, it is certainly suggestive of another component of gamma ray emission, such as that from annihilating dark matter. 

To explore furhter the possibility that a gamma ray component from dark matter annihilation is also present, we add an additional spherically symmetric template to our fit with a profile of emission proportional to $r^{-2 \gamma}$, where $\gamma$ is treated as a free parameter (if associated with dark matter annihilation, this parameter $\gamma$ is the same $\gamma$ that defines the inner slope of the dark matter profile). Although we fit each energy bin independently, we assume that $\gamma$ does not vary with energy. We find that the overall fit can be significantly improved by the inclusion of such a component, with the best fits found for values in the range of $\gamma = 1.18$ to 1.33 (our overall $\chi^2$ was reduced by 84 with the addition of 18 parameters).\footnote{Ref.~\cite{newanalysis} also finds that their fit the gamma ray emission in the Inner Galaxy is significantly improved by the inclusion of a dark matter-like component.}

\begin{figure}[!htb]
\resizebox{18.4cm}{!}{\includegraphics{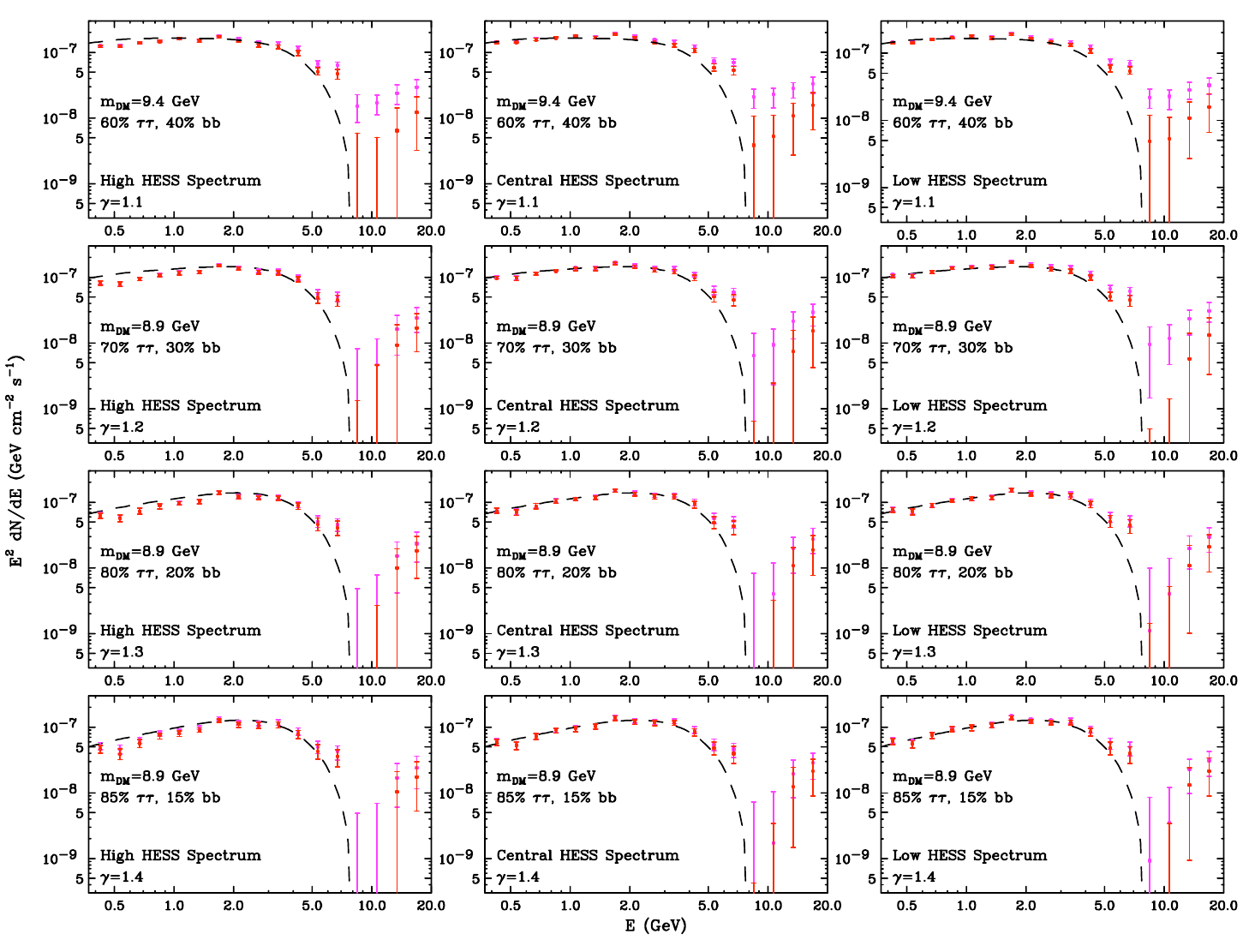}} \\
\caption{The spectrum of the spherically symmetric emission in the region within 1$^{\circ}$ of the Galaxy's dynamical center not associated with the disk, bulge, or resolved point sources, for each of three power-law expolations of the central HESS point source (see Fig.~\ref{spectracen}), and for several halo profile slopes. In each case, the spectrum shown has been corrected to account for the finite point spread function of FGST. The red (magneta) error bars have been (have not been) corrected for the ridge emission as observed by HESS (see text for details). In each frame, we compare the observed spectrum with that predicted from annihilating dark matter. Dark matter with a mass of 7-10 GeV, annihilating primarily to tau leptons can provide a good fit to the data.}
\label{spectradm}
\end{figure}

In Fig.~\ref{spectradm}, we show the spectrum of the additional spherically symmetric emission within 1$^{\circ}$ of the Milky Way's dynamical center, assuming either the central, upper, or lower power-law extrapolations for the central HESS source (as shown in Fig.~\ref{spectracen}). Results are also shown for several values of $\gamma$. In each frame, the red (magneta) error bars have been (have not been) corrected for the ridge emission as observed by HESS. If the ridge emission is accounted for, we find no statistically significant evidence for additional emission above $\sim$8 GeV.

This additional emission described here is both morphologically and spectrally consistent with that originating from annihilating dark matter. In each frame of Fig.~\ref{spectradm}, we show an example of a dark matter mass and annihilation channels that provides a good fit to the extracted spectrum. In general, we can accommodate the observed emission with dark matter particles with masses in the range of 7-10 GeV, and that annihilate primarily to $\tau^+ \tau^-$, but to hadronic channels 15-40\% of the time.


\section{Annihilating Dark Matter}
\label{dm}

In this section, we discuss the properties of the dark matter that are required to accommodate the excess gamma ray emission described in the prior section. As previously stated, good fits are found for dark matter masses in the range of approximately 7 to 10 GeV, and which annihilate primarily to a $\tau^+ \tau^-$ final state (in addition to any annihilations to final states which do not yield large fluxes of gamma rays, such as $e^+ e^-$, $\mu^+ \mu^-$, $\nu \bar{\nu}$, etc.), along with a smaller but not insignificant fraction to hadronic final states.

The value of the dark matter's annihilation cross section required to normalize the gamma ray spectrum to that shown in Fig.~\ref{spectradm} depends somewhat on how the distribution dark matter in the Inner Galaxy is itself normalized. The somewhat steep inner slope of the profile ($\gamma=1.18$ to 1.33) that is inferred from the data may be the result of adiabatic contraction~\cite{ac} occurring in the inner kiloparsecs of the Milky Way. If this is the case, then the profile's slope may be shallower ($\gamma \sim 1$) farther away the Galactic Center, for example.  

If we simply assume that a slope of 1.18 (1.33) continues to at least the Solar Circle, we find a value of $J$ (see Eqs.~\ref{flux2}-\ref{jpsi}) averaged over a radius of 1$^{\circ}$ around the Galactic Center of $4.1 \times 10^3$ ($1.9\times 10^4$), adopting a total mass density within the Solar Circle fixed to a that of an NFW profile with a local density of 0.3 GeV/cm$^3$. In contrast, if an NFW-like slope of $\gamma=1$ is found outside of the innermost 500 pc of Galaxy, at which point the transition to a slope of 1.18 (1.33) takes place, we find an averaged value of $J = 3.4 \times 10^3$ ($6.5 \times 10^3$). For the case of a dark matter particle with a mass of 7-10 GeV and that annihilates to $\tau^+ \tau^-$ and hadronic channels (as shown in Fig.~\ref{spectradm}), these profiles would imply an annihilation cross section in the range of $\sigv \approx 4.6 \times 10^{-27}$ cm$^3$/s to $5.3 \times 10^{-26}$ cm$^3$/s. If in addition to tau leptons, the dark matter also annihilates with an equal cross section to electrons and muons, a total cross section of $\sigv \approx 1.2 \times 10^{-26}$ cm$^3$/s to $1.3 \times 10^{-25}$ cm$^3$/s is required.

\section{Other Interpretations}
\label{other}

In this section, we discuss explainations other than dark matter annihilations that might account for the observed emission from the inner volume of the Milky Way. The spectral shape and angular distribution of the observed signal, however, are somewhat difficult to account for with known astrophysical sources or mechanisms. In particular, within the inner $\sim$$0.25^{\circ}$, the additional ({\it ie.} non-pion, non-inverse Compton) component dominates over the other sources of 1-5 GeV gamma rays, but is entirely indiscernible beyond $1.25^{\circ}$ from the Galactic Center. This requires the sources of the emission to be highly concentrated within a region only a few tens of parsecs in radius (recall that the best fit distribution of the sources of the excess emission is proportional to approximately $r^{-2.5}$). 

To account for the observed excess emission, one could consider a spatially concentated population of unresolved gamma ray point sources with spectra that peak in the 1-4 GeV range. Millisecond pulsars, with emission that is observed to fall off rapidly above a few GeV, represent such a possibility~\cite{kev}. In particular, based on observations of resolved millisecond pulsars in other regions of the sky by FGST, the average spectrum of such objects takes the form of $dN_\gamma/dE_{\gamma} \propto E_{\gamma}^{-1.5} \exp(-E_{\gamma}/2.8\,{\rm GeV})$~\cite{averagepulsar}. The spectrum of the excess emission observed from the Inner Galaxy, in contrast, rises with a somewhat harder spectral index, $dN_\gamma/dE_{\gamma} \propto E_{\gamma}^{-\Gamma} \exp(-E_{\gamma}/E_{\rm cut}\,{\rm GeV})$, with $\Gamma = 0.99^{+0.10}_{-0.09}$ and $E_{\rm cut}=1.92^{+0.21}_{-0.17}$ GeV. Among the 46 gamma ray pulsars (millisecond and otherwise) in the FGST's first pulsar catalog, the distribution of spectral indices peak strongly at $\Gamma=$1.4, with 44 out of 46 of the observed pulsars possessing (central values of their) spectral indices greater than 1.0~\cite{pulsarcatalog}. As a large number of pulsars would be required to generate the emission observed from the Inner Galaxy, their overall spectrum would naively be expected to be similar to that of the average observed pulsar. In order for the observed emission to result from pulsars, the population of pulsars present in the Inner Galaxy would have to differ significantly from the sample represented by the Fermi pulsar catalog. As the star forming environment of the Inner Galaxy is different from those found elsewhere throughout the Galaxy, this possibility cannot be discarded. An opportunity to measure the emission from large populations of gamma ray pulsars can be found in globular clusters, whose gamma ray emission is generally attributed to pulsars contained within their volumes. Unfortunately, the spectral indices of these objects are not currently well measured. In particular, the eight globular cluster spectra reported by Fermi have an average spectral index very close to that of pulsars ($\Gamma \approx 1.4$), but with individual error bars which extend from roughly 0 to 2.5. Perhaps with more data, we will learn from these systems whether the spectral indices of large pulsar populations can be hard enough to accommodate the emission observed from the Galactic Center. Lastly, we mention the morphological challenge faced by a pulsar interpretation of the emission from the Inner Galaxy. Even if we ignore the issue of the spectral shape of gamma ray pulsars, it is somewhat difficult to explain why the number density of such sources would increase so rapidly in the innermost tens of parsec of the Milky Way. Even modest pulsar kicks of $\sim 100$ km/s would allow a pulsar 10 pc from the Galactic Center to escape the region, consequently broading the angular width of the signal. Annihilating dark matter, in contrast, produces a flux of gamma rays that scales with its density {\it squared}, and thus can much more easily account for the high concentration of the observed signal.

Alternatively, one could consider the possibility that the excess emission originates from neutral pions produced in the collisions of cosmic rays with gas, but with a much harder spectrum than is found in other regions of the Galaxy. This, however, turns out to be implausible upon closer inspection.  If we harden the cosmic ray proton spectrum, the gamma ray peak does shift to the right, but not with the climb and sudden fall manifest in the spectrum observed from the Inner Galaxy. To generate such a peak at 2-3 GeV, one could consider a strongly broken power-law for the cosmic ray spectrum, but even this is not easily able to generate the observed gamma ray spectrum. To produce the observed shape, we would be forced to imagine a very extreme cosmic ray spectrum such as that given by $dN_p/dE_p \propto E^{-1}_p$ below 50 GeV, and suddenly dropping, for example as $dN_p/dE_p \propto E^{-5}_p$, above 50 GeV.


Lastly, we note that if at GeV energies the spectrum of the gamma ray emission from the Milky Way's supermassive black hole strongly departs from its power-law form observed by HESS, then it could account for much of the emission being discussed here. Although we find that the quality of the fit does significantly improve when a gamma ray component from annihilating dark matter is included (demonstrating that not all of the emission in question originates from a single point source), we leave open the possibility that this could be an artifact of the modelling of astrophysical backgrounds. As the measurement of the spectrum and morphology of the gamma ray emission from this region improves, it will become increasingly possible to discriminate between these possibilities. Of the astrophysical explanations proposed for the gamma ray emission observed from the Inner Galaxy, we consider this to be the most plausible.


\section{Comparisons With Our Previous Study}
\label{previous}

In a previous paper~\cite{lisa}, we analyzed the first year of FGST data from the inner $3^{\circ}$ around the Galactic Center, and reached similar (although not identical) conclusions to those described here. In particular, both studies identified a bump-like feature peaking at 2-5 GeV in the inner degree or so of the Inner Galaxy. We will now briefly discuss the differences between the results presented in Ref.~\cite{lisa} and those presented here.

For this work, we modelled the backgrounds in a significantly more sophisticated way than was done in Ref.~\cite{lisa}. Although Ref.~\cite{lisa} used a background with a similar morphology to our disk-component, spherically symmetric, bulge-like backgrounds were not considered. As a result, the slope of the best fit halo profile was found to be somewhat shallower than those found here ($\gamma \approx 1.1$ rather than 1.18 to 1.33). In other words, the excess component described in Ref.~\cite{lisa} consists of a combination of the excess component described in this paper and emission from pion decay (and inverse Compton scattering) in the bulge. As a consequence, the fits of Ref.~\cite{lisa} required a somewhat higher overall dark matter annihilation cross section that those found here.

Furthermore, the spectral shape of the background used in Ref.~\cite{lisa} was fit to a power-law, and thus did not account for the details of the pion decay (plus inverse Compton scattering) spectrum that was shown in this paper to well describe the observed disk and bulge emission. This had the net effect of skewing the spectral shape of the extracted excess in the inner degree (compare Fig.~2 of Ref.~\cite{lisa} to Fig.~\ref{spectradm} of this paper). The somewhat broader spectral feature found in Ref.~\cite{lisa} was shown to be adequately fit by a 25-30 GeV dark matter annihilating to $b\bar{b}$. This scenario, however, does not provide a good fit to the spectrum found here (see Fig.~\ref{spectradm}). Instead, we find good fits only for dark matter with a mass of 7-10 GeV, and that annihilates primarily to tau leptons.

We also note that the Fermi Collaboration has presented preliminary results from their analysis of the Galactic Center~\cite{prelim}. Although no concerete conclusions were reached, they do confirm the presence of a statistically significant excess (relative to astrophysical backgrounds) between approximatley 1.5 and 4 GeV, roughly consistent with the characteristics originally presented in Ref.~\cite{lisa} and as presented here.

\section{Discussion and Conclusions}
\label{discussion}

In this paper, we analyzed the first two years of data from the Fermi Gamma Ray Space Telescope (FGST) with the intention of constraining or identifying the products of dark matter annihilations taking place in the inner volume of the Milky Way. We have identified a component of gamma ray emission concentrated within the inner degree of the Galactic Center, with a spectrum peaking at 1-4 GeV (in $E^2$ units). This emission does not appear to be consistent with that anticipated from known astrophysical backgrounds (although much of the observed emission could potentially originate from the Milky Way's supermassibe black hole if its spectrum strongly departs from the power-law behavior observed by higher energies). In contrast, this signal can easily be accounted for by annihilating dark matter. In particular, the observed morphology and spectrum of the signal can be well fit by dark mater distributed in a profile described by $\rho\propto r^{-\gamma}$, $\gamma=1.18$ to 1.33, with a mass of 7 to 10 GeV, and that annihilates primarily to tau leptons (possibily in addition to other leptonic channels). Depending on how the dark matter distribution is normalized, the required annihilation cross section falls within the range of $\sigv = 4.6 \times 10^{-27}$ to $5.3\times 10^{-26}$ cm$^3$/s (or $1.2 \times 10^{-26}$ to $1.3\times 10^{-25}$ cm$^3$/s if the dark matter annihilates to muons and electrons as often as it does to taus).

The characteristics of the dark matter implied by these observations are consistent with theoretical expectations. In particular, the required dark matter distribution is in concordance with that predicted by numerical simulations~\cite{nfw}, after accounting for a modest degree of adiabatic contraction~\cite{ac}. Furthermore, the annihilation cross section required to normalize the signal is consistent with that required of a simple thermal relic that freezes out with an abundance equal to the measured cosmological density of dark matter. The dark matter mass implied by this observation is also remarkably similar to that needed to simultaneously explain the observations by the direct detection experiments CoGeNT~\cite{cogent} and DAMA~\cite{dama}. In particular, it was shown in Ref.~\cite{consistent} that the signals reported by the CoGeNT and DAMA collaborations can be consistently interpreted as elastically scattering dark matter particles with masses in the range of approximately 6.2 to 8.6 GeV (see also Ref.~\cite{theory}).

In the dark matter scenario implied by the observations described in this paper, gamma rays are not the only potentially observable annihilation products. In particular, the electrons produced in the annihilations of dark matter (through either subsequent tau decays, or through annihilations directly to $e^+ e^-$ and/or $\mu^+ \mu^-$) will produce synchrotron emission peaking at frequences of $\nu_{\rm syn} \sim 23~{\rm GHz}\,\times \, (E_e/7~{\rm GeV})^2 \, (B/100 \,\mu{\rm G})$. Given the recently reported evidence of 100 microGauss-scale magnetic fields in the Inner Galaxy~\cite{nature}, it is not difficult to interpret the so-called WMAP haze~\cite{haze} as a signal of the dark matter particle described in this paper~\cite{linden}.

\bigskip
\section*{Acknowledgments}

We would like to thank Greg Dobler for providing the curves presented in Fig.~\ref{galprop}. We would also like to thank Scott Dodelson, Doug Finkbeiner, Patrick Fox, Joachim Kopp, Tim Linden, Tracy Slatyer, and Kathryn Zurek for helpful discussions and comments. We would also like to thank Alexey Boyarsky, Denys Malyshev, Oleg Ruchayskiy, who have recently studied the same data discussed here~\cite{newanalysis}. Discussions with them have been especially fruitful. This work has been supported by the US Department of Energy and by NASA grant NAG5-10842. Fermilab is operated by Fermi Research Alliance, LLC under Contract No.~DE-AC02-07CH11359 with the United States Department of Energy.

\end{document}